\documentclass[twoside,11pt]{article}

\PassOptionsToPackage{dvipsnames,svgnames,x11names}{xcolor}
\usepackage[preprint,nohyperref]{jmlr2e}

\usepackage[T1]{fontenc}
\usepackage{lmodern}
\usepackage{amsmath,amsfonts}

\usepackage{array,makecell,booktabs}
\newcolumntype{M}[1]{>{\centering\arraybackslash}m{#1}}

\usepackage{xcolor,tikz}
\usetikzlibrary{positioning,arrows.meta}
\usepackage{float}

\usepackage{enumitem}
\usepackage{algorithm,algpseudocode}

\usepackage[section]{placeins}
\usepackage{soul}

\DeclareMathOperator{\Cov}{Cov}
\DeclareMathOperator{\Var}{Var}
\DeclareMathOperator{\Corr}{Corr}

\usepackage{xurl}
\usepackage[hidelinks]{hyperref}
\hypersetup{breaklinks=true}
\usepackage{cleveref}

\newtheorem{prop}{Proposition}
\newtheorem{lem}{Lemma}
\newtheorem{cor}{Corollary}

\crefname{prop}{Proposition}{Propositions}
\crefname{lem}{Lemma}{Lemmas}
\crefname{cor}{Corollary}{Corollaries}
\crefname{rem}{Remark}{Remarks}

\renewenvironment{proof}[1][Proof]
  {\par\noindent\textbf{#1}\ }
  {\hfill\BlackBox\par\medskip}

\allowdisplaybreaks

\usepackage{lastpage}
\ShortHeadings
  {Environment-Adaptive Covariate Selection}
  {Zuo and Wang}
\firstpageno{1}

\begin{document}

\title{Environment-Adaptive Covariate Selection: \\Learning When to Use Spurious Correlations for Out-of-Distribution~Prediction}

\author{\name Shuozhi Zuo \email shuozhi@umich.edu \\
  \addr Department of Statistics\\
       University of Michigan\\
       Ann Arbor, MI 48109, USA
  \AND
  \name Yixin Wang \email yixinw@umich.edu \\
  \addr Department of Statistics\\
       University of Michigan\\
       Ann Arbor, MI 48109, USA
}

\editor{}

\maketitle

\begin{abstract}
A common approach to out-of-distribution prediction restricts models to causal or invariant covariates to avoid spurious associations that may change across environments. Despite its theoretical appeal, this strategy can underperform empirical risk minimization when only a subset of the causal parents of the outcome is observed. In such settings, non-causal covariates can serve as proxies for unobserved causal parents and improve prediction when the proxy relationship is stable, but they can hurt when shifts disrupt that relationship. Thus, the optimal covariate set can depend on the specific shift encountered. Because different shifts can leave distinct signatures in the covariate distribution, we propose an environment-adaptive covariate selection algorithm that chooses which covariates to use in each environment based on summaries of its unlabeled covariates. The summaries may be hand-crafted or learned from multi-environment data, and prior causal knowledge can be incorporated as constraints. We give sufficient conditions under which such summaries are informative in linear causal models, and explain when adapting only the covariate set is preferable to letting every coefficient vary with the environment. In simulations and two applications, adapting the covariate set improves prediction over static baselines, including causal and invariant covariate selection.
\end{abstract}

\section{Introduction}

A common approach to out-of-distribution (OOD) prediction restricts models to causal or invariant covariates, to avoid spurious, non-causal associations that may be unstable across environments. The rationale is that when distributional shifts preserve the mechanism that generates the outcome, the conditional distribution of the outcome given its causal parents remains unchanged, whereas the outcome's associations with non-causal covariates need not \citep{Pearl2009causality, peters2017elements, Scholkopf2021towards}. A large body of work builds on this reasoning and uses causal or invariant structure as a route to robustness under distribution shift \citep{peters2016causal, RojasCarulla2018, arjovsky2019invariant, Buhlmann2020invariance, Rothenhausler2021anchor}.

Despite its strong theoretical appeal, this strategy often falls short in practice: predictors restricted to causal or invariant covariates frequently fail to outperform empirical risk minimization (ERM) \citep{vapnik1999nature}, which uses all available covariates, and can perform worse, even when test environments differ substantially from training environments \citep{nastl2024do, gardner2023tableshift, Salaudeen2025domain}. This gap between theory and practice motivates the following question: \emph{when and why does causal-only covariate selection underperform for OOD prediction?}

\begin{figure}[t]
\centering
\includegraphics[width=\linewidth]{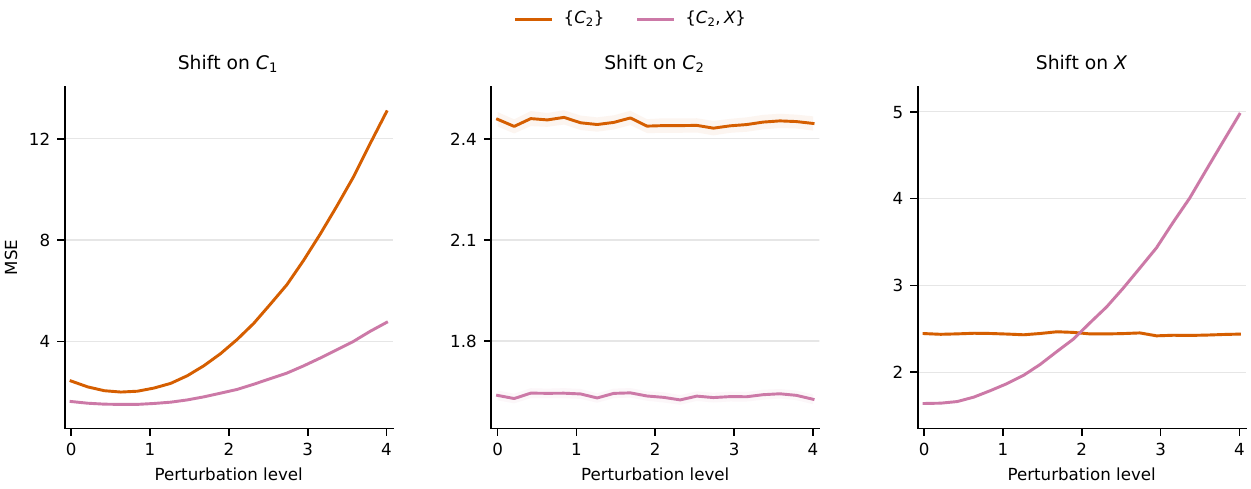}
\caption{Mean squared error (MSE) of the predictors using $\{C_2\}$ and $\{C_2, X\}$ under perturbations of $C_1$, $C_2$, and $X$. Shaded regions show pointwise 95\% confidence intervals over 1{,}000 replications. Including the proxy improves prediction while the proxy relationship is stable, but degrades it under sufficiently strong perturbations of $X$.\label{fig:sim}}
\end{figure}

We show that such failures can arise when only a subset of the causal parents of the outcome is observed. Observed causal covariates then miss part of the predictive signal, while non-causal covariates can serve as proxies for the unobserved parents. In the running example in \Cref{fig:dag_perturbations}, $Y$ depends on an observed parent $C_2$ and an unobserved parent $C_1$, and $X$ proxies $C_1$. Including $X$ improves prediction when this proxy relationship is stable, but can degrade performance under sufficiently strong direct perturbations to $X$~(\Cref{fig:sim}).

\begin{figure}[t]
\centering
\definecolor{EACSBlue}{HTML}{0072B2}
\definecolor{EACSInk}{HTML}{333333}
\definecolor{EACSHidden}{HTML}{E6E6E6}
\begin{tikzpicture}[
    var/.style={circle, draw=EACSInk, line width=0.55pt, fill=white,
      minimum size=7.2mm, inner sep=0pt},
    hidden/.style={var, fill=EACSHidden},
    edge/.style={-{Stealth[length=1.7mm, width=1.15mm]}, draw=EACSInk, line width=0.65pt},
    perturb/.style={-{Stealth[length=2mm, width=1.4mm]}, draw=EACSBlue, line width=1.1pt},
    title/.style={font=\sffamily\small, anchor=base}
]
\foreach \dx/\t/\heading in {0/a/{Causal graph}, 4.9/b/{Perturbation on $C_1$}, 9.8/c/{Perturbation on $X$}} {
  \begin{scope}[xshift=\dx cm]
    \node[title] at (1.4, 1.82) {(\t) \heading};
    \node[hidden] (C1\t) at (1.4, 0.78)  {$C_1$};
    \node[var]    (C2\t) at (1.4, -0.78) {$C_2$};
    \node[var]    (X\t)  at (0, 0)       {$X$};
    \node[var]    (Y\t)  at (2.8, 0)     {$Y$};
    \foreach \a/\b in {C1/X, C1/Y, C2/X, C2/Y}
      \draw[edge] (\a\t) -- (\b\t);
  \end{scope}
}
\draw[EACSBlue, line width=1.1pt] (C1b) circle (3.6mm);
\draw[perturb] ([yshift=4mm]C1b.north) -- (C1b.north);
\draw[EACSBlue, line width=1.1pt] (Xc) circle (3.6mm);
\draw[perturb] ([yshift=4mm]Xc.north) -- (Xc.north);
\end{tikzpicture}
\caption{(a) Causal graph of the running example: the outcome $Y$ has an unobserved causal parent $C_1$ (shaded) and an observed causal parent $C_2$, and the covariate $X$ depends on both and acts as a proxy for $C_1$. (b--c) Perturbations of the unobserved parent $C_1$ and of the proxy $X$, marked by blue outlines and arrows.\label{fig:dag_perturbations}}
\end{figure}

The optimal set of predictive covariates is therefore not universal: it depends on the type of shift.

\begin{figure}[t]
    \centering
    \includegraphics[width=\linewidth]{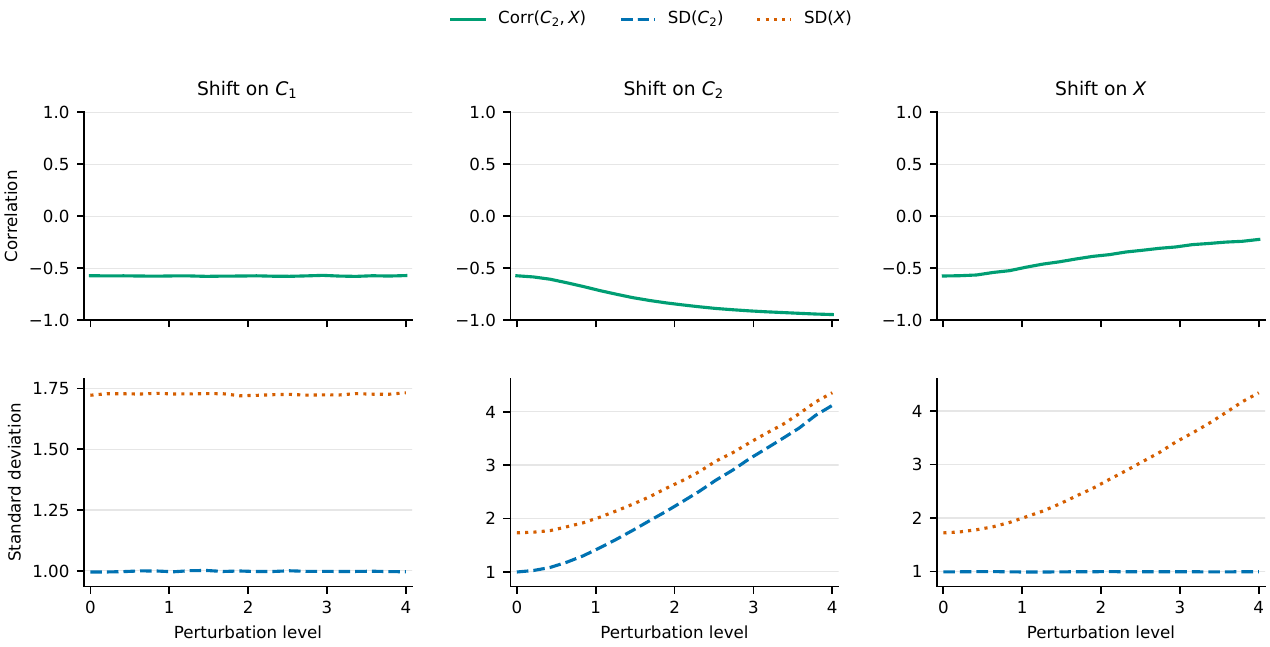}
\caption{Summary statistics of the covariate distribution across environments, corresponding to the perturbation scenarios in \Cref{fig:sim}. Different types of distribution shift induce distinct, observable signatures, such as changes in variance and dependence structure. These summaries can be computed from unlabeled covariates in the target environment and used to assess whether the proxy covariate remains reliable.\label{fig:sim_a_sum}}
\end{figure}

Although the shift type is not directly observed, different shifts can leave \emph{observable signatures} in the covariate distribution. For example, perturbations that add noise to a proxy covariate alter its variance and its correlation with observed causal parents, whereas mean shifts in an unobserved parent leave a different signature. As shown in \Cref{fig:sim_a_sum}, simple summaries of the unlabeled target covariates can distinguish environments in which $X$ is helpful from those in which it becomes harmful. 

Motivated by these observations, we propose \emph{environment-adaptive covariate selection} (EACS). Rather than committing to a static covariate subset, EACS uses multiple training environments to learn a rule that maps summaries of the covariate distribution to the covariates used for prediction, to minimize prediction risk in each environment. These summaries may be simple, such as variances or correlations in the running example, or learned directly from multi-environment covariate data when the relevant distributional features are not known in advance. At test time, EACS computes the summary from the unlabeled target covariates and chooses the covariates for that environment accordingly. Because the predictor for each covariate subset is fitted on all pooled observations, the training environments are needed only to learn when to keep or drop proxies, not how every coefficient varies across environments. Prior causal knowledge can also be incorporated as constraints.

\begin{figure}[t]
    \centering
    \includegraphics[width=\linewidth]{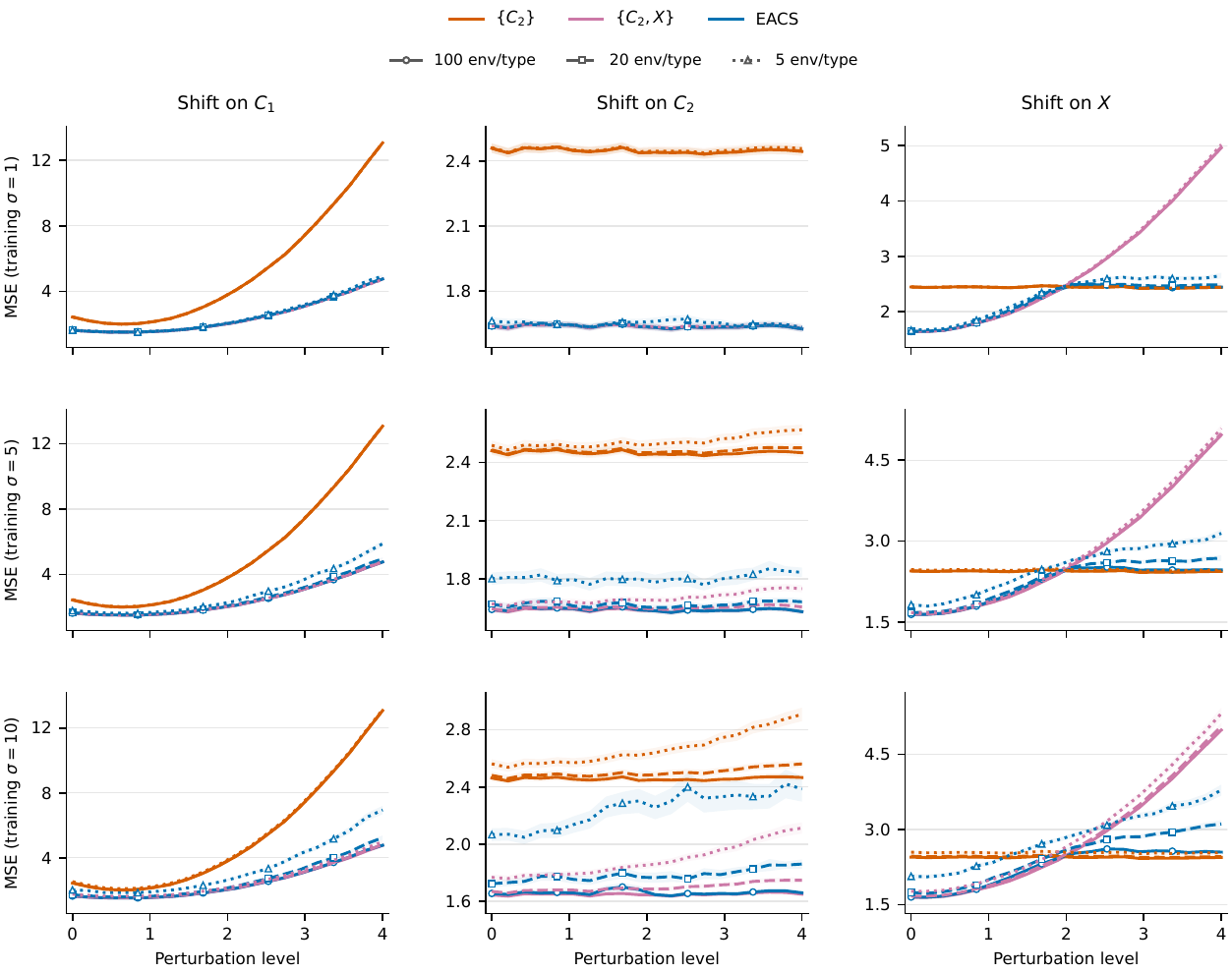}
\caption{MSE of EACS and of the fixed predictors using $\{C_2\}$ or $\{C_2,X\}$ in the running example, for three shift types (columns) and three training outcome noise levels (rows). Colors identify predictors; line styles and markers identify the number of training environments per shift type. Shaded regions show pointwise 95\% confidence intervals over 1{,}000 replications.\label{fig:sim_a_env_mse}}
\end{figure}

In the running example, EACS keeps the proxy where it helps and drops it where it hurts (\Cref{fig:sim_a_env_mse}). In simulations and two applications, it improves prediction over static baselines, including causal and invariant covariate selection. The results suggest that robust OOD prediction need not avoid spurious covariates altogether; it can instead learn when such covariates can be trusted.

\noindent\textbf{Contributions.} The contributions of this paper are as follows:
\begin{enumerate}[label=(\alph*), noitemsep, topsep=0pt, leftmargin=*]
\item Building on results that causal predictors can be suboptimal under shifts of bounded strength \citep{Rothenhausler2021anchor, christiansen2022causal} and that unobserved causes can break invariance \citep{christiansen2020switching}, we analyze proxies of unobserved causal parents as a mechanism by which causal and invariant covariate selection can underperform for OOD prediction: non-causal covariates may proxy the missing parents, and excluding them can reduce predictive accuracy.
\item We show that proxy usefulness is shift-dependent. In linear structural causal models, we give sufficient conditions under which summaries of the unlabeled covariate distribution reveal whether proxies remain reliable, and show that no such summary suffices when shifts on an unobserved parent and on its proxy leave the same signature in the covariates but favor different subsets.
\item We propose EACS, which learns to select covariates from environment summaries, either hand-crafted or learned from multi-environment data. We give a finite-sample bound on the risk of the learned selector relative to the best selector in its class, and show empirical gains over static baselines, including causal and invariant selection, in simulations and two applications, although the gains in the census income application are small. We also show in a controlled simulation when restricting adaptation to the choice of covariates is preferable to contextual prediction (which lets every coefficient vary with the environment): when many covariates need no adaptation or the training shifts do not cover the test shifts.
\end{enumerate}

\noindent\textbf{Related work.} This work draws on several lines of research on OOD prediction and distribution shift.

\ul{Causal and invariant prediction.} \citet{peters2016causal} proposed invariant causal prediction, which uses invariance across environments to infer causal parents of the outcome with confidence guarantees. Subsequent work extended invariant prediction to nonlinear models \citep{heinze2018invariant} and to representation learning via invariant risk minimization \citep{arjovsky2019invariant}, which seeks representations whose optimal predictors are invariant across environments. \citet{Fan2024environment} develop environment-invariant linear least squares, a regularized approach for estimating invariant parameters across environments with guarantees for consistent estimation and variable selection. \citet{Wu2025bayesian} propose a Bayesian formulation that treats invariance as latent structure and enables posterior inference in high-dimensional settings. Causal predictors are worst-case optimal when the class of shifts is sufficiently rich \citep{RojasCarulla2018, christiansen2022causal}, but under shifts of bounded strength other predictors can be more accurate \citep{Rothenhausler2021anchor}. Empirically, predictors restricted to causal features or trained with invariance-based objectives often fail to outperform ERM \citep{nastl2024do, gardner2023tableshift}, and \citet{Salaudeen2025domain} argue that many benchmarks favor ERM because their shifts do not disrupt the spurious correlations it exploits.

\ul{Unobserved causes and proxies.} Several works relax strict invariance or address unobserved causal variables directly. \citet{Rothenhausler2021anchor} introduced anchor regression, which uses external anchor variables to navigate the trade-off between robustness and predictive accuracy. Building on this idea, \citet{oberst2021regularizing} study linear prediction when the anchor variables are unobserved but noisy proxies are available, using proxy-based regularization to trade off in-distribution performance and robustness to interventions on the unobserved variables. They also consider a targeted setting, in which the mean and covariance of the test-time shift are known and used to tailor the predictor to that shift. \citet{christiansen2020switching} extend invariance-based causal discovery to settings in which some discrete direct causes of the outcome are unobserved, so that the outcome given its observed causal parents need not be invariant. \citet{mazaheri2023causal} select and engineer features from observed descendants of the unobserved causes and effects of the outcome so that a single predictor remains stable across environments. Proxy-based domain adaptation uses proxies of a latent variable whose distribution shifts, together with unlabeled target data, to adapt the predictor to the target domain \citep{alabdulmohsin2023adapting, tsai2024proxy}.

This paper shares with \citet{mazaheri2023causal} the premise that proxies of unobserved causes carry useful but shift-sensitive information, and with the targeted setting of \citet{oberst2021regularizing} and proxy-based adaptation the idea of using information about the target environment to move beyond invariance. The difference lies in how that information is used. The targeted approach adapts a linear predictor to specified moments of the anchor shift, proxy-based adaptation assumes that the shift acts only through the latent variable and so leaves the proxy relationships intact, and \citet{mazaheri2023causal} construct a single stable predictor. EACS instead learns from the training environments which features of the unlabeled covariate distribution indicate that proxies remain reliable, and selects an environment-specific covariate subset accordingly; it does not require anchors or the moments of the shift to be specified, but it needs training environments in which some shifts disrupt the proxy relationship.

\ul{Covariate shift, reweighting, and robust optimization.} Classical covariate-shift methods assume that the input distribution changes between training and test populations while the conditional output distribution remains stable, and use density-ratio weights to target a fixed test population \citep{shimodaira2000improving, sugiyama2007direct}. EACS addresses a complementary regime: when some causal parents are unobserved, the observed relationship between the outcome and proxy covariates can vary across environments even if the structural equation for the outcome is unchanged. Rather than reweighting the training sample toward the target covariate distribution, EACS uses the target covariates to infer whether proxy covariates remain reliable.

Distributionally robust optimization takes a different route, learning predictors with low worst-case risk over a prespecified uncertainty set. \citet{duchi2020distributionally} study worst-case performance over latent covariate mixtures, and \citet{sahoo2022learning} learn predictors that are robust to sample-selection bias in which unobserved factors have a bounded effect on the selection probability. These methods typically return a single minimax predictor for a chosen robustness set, whereas EACS adapts to each test environment, using its covariate distribution to select among covariate subsets.

\ul{Adaptation using unlabeled target data.} Another line of work investigates how information from the target environment can be used to improve OOD generalization, particularly when outcome labels are unavailable. Domain generalization by marginal transfer learning \citep{blanchard2011generalizing, blanchard2021domain} augments each observation with an embedding of the marginal covariate distribution of its domain and learns a single predictor that takes both as input, with generalization bounds that, like ours, depend on both the number of training domains and the number of observations per domain. Adaptive risk minimization learns across training domains how to adapt predictions using unlabeled target examples \citep{zhang2021adaptive}. In-context risk minimization \citep{gupta2024context} trains a sequence model that conditions on unlabeled examples from the same environment, so that its predictions approach the risk minimizer of the test environment. \citet{jeong2025out} reweight the training environments to match covariate moments of the target environment. Closer to our aim of using spurious features when they help, \citet{eastwood2023spuriosity} adapt how unstable features are used in each test domain with pseudo-labels from stable features, and \citet{sun2023beyond} reweight a source classifier to track test-time changes in the joint distribution of labels and spurious attributes. EACS differs in restricting the adaptation: rather than adapting the predictor itself, it adapts only which covariates are used and shares the predictors across environments. This restriction keeps the selection interpretable and allows causal constraints. \citet{magliacane2018domain} also adapt only the choice of covariates, using conditional independence tests on labeled source data and unlabeled target data to select a subset whose conditional relationship with the outcome is the same in both domains, allowing for hidden confounders; EACS does not require this invariance and can retain proxies whose relationship with the outcome changes but remains informative. We use causal structure to explain when the target covariate distribution carries the information needed for this choice, and we show when the restricted adaptation is preferable to contextual prediction (\Cref{rem:env_specific,sec:contextual_main}).

\ul{Method and model selection under distribution shift.} A further line of work emphasizes that no single algorithm performs best across all types of distribution shift and studies how to select methods to achieve robustness under varying shifts. \citet{bell2024reassessing} show that spurious-correlation benchmarks often disagree about which methods perform best, and suggest choosing methods using the benchmark most similar to the problem at hand. \citet{jiang2025ood} train a selector on a collection of datasets with controlled shifts to choose a training algorithm from characteristics of a dataset and its shift. Relatedly, \citet{deng2024autoeval} predict the accuracy of a fixed classifier on an unlabeled test set by regressing accuracy on distribution statistics across many synthetically shifted datasets. EACS similarly learns across environments how covariate summaries relate to predictive performance, but uses this relation to choose covariates for each test environment rather than to choose a training algorithm or to evaluate a fixed model.

\noindent\textbf{Organization.} Section~2 reviews causal and invariant prediction and uses the running example to show why no fixed covariate subset is uniformly optimal. Section~3 introduces EACS and its discrete and continuous implementations, and shows how to incorporate prior causal knowledge. Section~4 gives theoretical guarantees, explains why EACS restricts adaptation to the choice of covariates, and examines when environment summaries are informative. Section~5 presents the empirical studies, and Section~6 discusses limitations and future work.

\section{When Does Causal-Only Covariate Selection Underperform ERM in OOD Prediction?}

OOD prediction is simplest when all causal parents of the outcome are observed: ERM on the observed covariates can coincide with causal and invariant prediction. We study the setting in which only some causal parents are observed. Predictive accuracy can then degrade because part of the signal is missing, and invariance with respect to the observed causal parents can break down when shifts act on the unobserved parents, even though the causal mechanism for the outcome is unchanged.

We first revisit invariant and causal prediction, then return to the running example to show how different perturbations affect predictive performance.

\subsection{Invariant models for OOD prediction}

Invariance formalizes the stability of covariate-outcome relationships across environments, even when the distributions of the covariates themselves change. Given a collection of environments $\mathcal{E}$, a subset $V$ of the observed covariates is invariant \citep{peters2016causal} if, for all $e, e' \in \mathcal{E}$,
\[
\mathbb{P}_e(Y \mid V) = \mathbb{P}_{e'}(Y \mid V).
\]
Predictors based on such covariates remain reliable under distributional shifts that preserve this conditional distribution.

Restricting prediction to invariant relationships can be overly conservative, however, when only some causal parents are observed: important predictive signals may be discarded because their associations with the outcome are not invariant across environments. When these associations nonetheless remain stable under the shift actually encountered, invariant approaches often underperform ERM.

\subsection{Causal models for OOD prediction}

Causal prediction focuses on how interventions on covariates affect the outcome. Let $C$ denote the causal parents of $Y$. Using Pearl's \emph{do}-notation, the causal predictor is
\[
f_{\text{causal}}(c) = \mathbb{E}[Y \mid \mathrm{do}(C = c)].
\]
When shifts leave the mechanism that generates $Y$ unchanged and there is no unmeasured confounding, the conditional distribution $\mathbb{P}(Y \mid C)$ is invariant across environments, and $f_{\text{causal}}(c) = \mathbb{E}[Y \mid C = c]$. Under additive noise, the causal predictor also satisfies a minimax optimality property \citep{RojasCarulla2018}:
\[
f_{\text{causal}}
\in \arg\min_f \sup_{P \in \mathcal{P}}
\mathbb{E}_{P}\!\left[(Y - f(C, X))^2\right],
\]
where $f$ ranges over functions of all covariates, both the causal parents $C$ and the remaining covariates $X$, and $\mathcal{P}$ ranges over distributions arising from shifts that preserve the mechanism that generates $Y$.

In practice, however, some causal parents are often unobserved. Let the full set of causal parents be $C = (C_s, U)$, where $C_s$ are observed and $U$ are unobserved. The conditional distribution $\mathbb{P}(Y \mid C_s, U)$ remains invariant across environments, but the observable conditional becomes
\[
\mathbb{P}_e(Y \mid C_s) = \int \mathbb{P}(Y \mid C_s, U)\, \mathbb{P}_e(U \mid C_s)\, dU.
\]
Because $\mathbb{P}_e(U \mid C_s)$ may vary across environments, even an unchanged causal mechanism $\mathbb{P}(Y \mid C_s, U)$ can yield
\[
\mathbb{P}_e(Y \mid C_s) \neq \mathbb{P}_{e'}(Y \mid C_s)
\quad \text{for some } e, e' \in \mathcal{E}.
\]
Invariance with respect to the observed causal parents can therefore break down when perturbations change the conditional distribution of the unobserved causal parents given the observed ones. A related phenomenon is studied by \citet{christiansen2020switching}, where a discrete latent variable modulates the outcome mechanism and invariance given the observed parents fails.

\subsection{Proxy covariates, unobserved causal parents, and environment-dependent robustness}

When some causal parents of the outcome are unobserved, other covariates may serve as \emph{proxies} for the missing parents, carrying indirect information that partially recovers the lost predictive signal. This helps explain why ERM, which uses all available covariates, can perform well despite lacking formal invariance guarantees.

The benefit of a proxy depends on the shift. If its link to the unobserved parent remains stable, including the proxy improves prediction; if a shift disrupts that link, the proxy can become unreliable and harm performance. These trade-offs arise in settings such as healthcare, where laboratory measurements proxy latent physiological states, and socioeconomic applications, where survey responses proxy latent preferences or local conditions; in both cases, changes in measurement or reporting practices can weaken the proxy relationship.

No single covariate subset is therefore uniformly optimal when causal parents are partially unobserved.

\subsection{Running example: proxy covariates under different distribution shifts}
\label{section:running-example-init}

We now return to the running example to illustrate how proxy usefulness changes across shifts. As in \Cref{fig:dag_perturbations}, $Y$ depends on an observed parent $C_2$ and an unobserved parent $C_1$, with $X$ proxying $C_1$.

To quantify these effects, we simulate data from the linear Gaussian model
\[
Y = C_1 + C_2 + \varepsilon_Y, \qquad X = C_1 - C_2 + \varepsilon_X,
\]
with independent $C_1, C_2, \varepsilon_Y, \varepsilon_X \sim \mathcal{N}(0,1)$. Environments are generated by perturbing one variable at a time: shifting the mean of $C_1$ by $\delta$, or adding independent Gaussian noise $\mathcal{N}(0,\delta^2)$ to $C_2$ or to $X$, with $\delta \in [0,4]$. We compare a model that uses only the observed causal parent, $\{C_2\}$, with a model that uses both $\{C_2, X\}$. Both are fitted by pooled least squares on 20 training environments per shift type, each training and test environment contains 100 observations, and performance is measured by mean squared error (MSE) over 1{,}000 replications.

\Cref{fig:sim} summarizes the results. Both models are sensitive to shifts in $C_1$, but as long as the proxy remains reliable, including $X$ lowers the error by providing information about $C_1$. When $X$ is heavily perturbed, however, its signal deteriorates, and the model using only $C_2$ eventually outperforms the model that includes the proxy.

For this data-generating process, we can characterize analytically when the proxy $X$ helps or harms prediction. Consider the population pooled least-squares predictors fitted on the training environments, and let $\beta_3$ be the coefficient of $X$ in the predictor using $\{C_2, X\}$. Let $s_{2,e}$ and $s_{3,e}$ be the standard deviations (SDs) of $C_2$ and $X$ in environment $e$, let $\mu_e = \mathbb{E}_e(C_1)$, and let $\mu_0$ be its pooled training mean. Because $\mathrm{Var}_e(C_1)=1$ in every environment of this example, the environment-specific risk difference is
\begin{align*}
\Delta_e &:= R_e(\{C_2\}) - R_e(\{C_2, X\})\\
&= 2\beta_3 - \beta_3^2\,(s_{3,e}^2 - s_{2,e}^2) + \beta_3(2-\beta_3)(\mu_e-\mu_0)^2 .
\end{align*}
The causal-only subset $\{C_2\}$ therefore has lower risk ($\Delta_e < 0$) exactly when
\[
s_{3,e}^2 - s_{2,e}^2 > \frac{2 + (2-\beta_3)(\mu_e-\mu_0)^2}{\beta_3},
\]
that is, when $X$ is sufficiently noisy relative to $C_2$. Since $0<\beta_3<1$ (\Cref{sec:sim_analytic_rule}), the last term of $\Delta_e$, $\beta_3(2-\beta_3)(\mu_e-\mu_0)^2$, is nonnegative: mean shifts in the unobserved parent can only make the proxy more valuable. An equivalent condition in terms of the correlation $r_e = \Corr_e(C_2, X)$ is given in \Cref{sec:sim_analytic_rule}.

This condition yields clear predictions for the three shift types:
\begin{enumerate}[label=(\alph*)]
\item \emph{Mean shifts in $C_1$} do not alter $s_{2,e}$, $s_{3,e}$, or their difference $s_{3,e}^2 - s_{2,e}^2$; they change only the last term of $\Delta_e$, which is nonnegative and so never favors excluding $X$. The optimal covariate subset is therefore unchanged.
\item \emph{Adding noise to $C_2$} increases the variance $s_{2,e}^2$ and, through the relation $X = C_1 - C_2 + \varepsilon_X$, increases $s_{3,e}^2$ by the same amount. The difference $s_{3,e}^2 - s_{2,e}^2$ is therefore unchanged, and so is the optimal covariate subset.
\item \emph{Adding noise to $X$} increases $s_{3,e}^2$ without affecting $s_{2,e}^2$, raising $s_{3,e}^2 - s_{2,e}^2$ and potentially flipping the inequality, so the optimal covariate subset can change.
\end{enumerate}
These predictions match the empirical patterns: perturbing $C_1$ or $C_2$ leaves the risk ordering unchanged, whereas perturbing $X$ produces a crossover where excluding the proxy becomes advantageous.

\paragraph{What invariance-based methods do in this example.}
Invariance-based methods retain covariate subsets whose conditional relationship with $Y$ is stable across training environments. Under mean shifts of $C_1$, no subset of the observed covariates is invariant: $C_1$ affects $Y$ but is not observed, and $X$ reflects it only with noise, so the conditional mean of $Y$ given any subset of $(C_2, X)$ moves with $\mu_e$. For example, $\mathbb{E}_e(Y \mid C_2) = C_2 + \mu_e$ and $\mathbb{E}_e(Y \mid C_2, X) = (3C_2 + X + \mu_e)/2$. Under noise added to $C_2$ or to $X$, in contrast, $C_1$ is unaffected and $\{C_2\}$ is invariant. Thus, the observed causal parent is invariant only when the shifts leave the unobserved parent untouched. When the training environments include mean shifts of $C_1$, as in our simulations, an invariance-based method finds no invariant subset to rely on; falling back to the pooled mean then gives higher risk than the causal-only predictor $\{C_2\}$ in every environment. The invariance-based baseline in the bike-sharing study (\Cref{sub:Bike}) shows such a collapse to the intercept-only model.

\section{Environment-Adaptive Covariate Selection}
\label{sec:aefs}

We now introduce EACS, an algorithm for selecting predictive covariates that adapts to distribution shifts across environments. EACS learns a selector that maps a summary of the target covariate distribution to weights over a library of covariate-subset predictors, to minimize prediction error in the target environment. This allows the method to include proxy covariates when the target summary suggests that they remain reliable, and to exclude them when the summary suggests that the proxy relationship has been disrupted.

The section proceeds as follows. \Cref{subsec:causal_shift_class} defines the causal model and shift class. \Cref{subsec:risk_based,subsec:env_repr,subsec:learn_selector,subsec:prediction_interp} formalize risk-based selection, environment representations, the selector, and test-time prediction. \Cref{subsec:implementations} presents the discrete and continuous implementations, and \Cref{sec:causal-selector} shows how to incorporate prior causal knowledge. Notation is summarized in \Cref{tab:eacs_notation}.

\subsection{Causal model and shift class}
\label{subsec:causal_shift_class}

We model the data-generating process as a structural causal model over the outcome $Y$, its observed causal parents $C_s$, its unobserved causal parents $U$, and a vector of observed non-causal covariates $X$; we write $W = (C_s, X)$ for the observed covariates. We consider environments that differ only through interventions on the covariates $(C_s, U, X)$; the structural equation for $Y$,
\[
Y = f_Y(C_s, U, \varepsilon_Y),
\]
is therefore preserved across environments.

\paragraph{Shift class.}
Each environment $e$ is identified with an intervention $i(e)$ on the covariates $(C_s, U, X)$. We write $\mathcal{I}$ for the class of admissible interventions and $\mathcal{E} = \{e : i(e) \in \mathcal{I}\}$ for the resulting shift class; the shifts in the running example are members of $\mathcal{I}$.

\paragraph{The optimal subset is determined by the intervention.}
Although $f_Y$ is preserved across $\mathcal{E}$, interventions on the covariates can alter the proxy relationship between $X$ and the unobserved causal parent $U$: an intervention that preserves this relationship favors including $X$, and one that disrupts it strongly enough favors causal-only selection. The optimal subset also depends on the training environments, because the subset predictors $f_z$ of \Cref{subsec:risk_based} are fitted on them. Given the fitted predictors, the risk of each candidate subset in environment $e$ is determined by the intervention, so the optimal subset $z^\star(e)$ defined in \Cref{subsec:risk_based} satisfies
\[
z^\star(e) = z^\star\!\big(i(e)\big).
\]

\paragraph{Knowledge of the causal parents is not required.}
The partition of $W$ into $C_s$ and $X$ is used to describe the mechanism, not as an input to the algorithm. EACS selects among subsets of all observed covariates and does not need to know which of them are causal parents of $Y$. When such knowledge is available, it can be imposed as a constraint on the candidate subsets (\Cref{sec:causal-selector}).

\begin{table}[t]
\centering
\begin{tabular}{ll}
\toprule
\textbf{Symbol} & \textbf{Description} \\
\midrule
$\mathcal{I}$ 
& Class of admissible interventions on the covariates \\
$\mathcal{E}$ 
& Shift class: environments generated by interventions in $\mathcal{I}$ \\
$i(e) \in \mathcal{I}$ 
& Intervention identifying environment $e \in \mathcal{E}$ \\
$W=(C_s,X)$
& Full observed covariate vector\\
$Z \subseteq \{0,1\}^p$ 
& Finite library of candidate subsets of coordinates of $W$ \\
$z \in Z$ 
& Binary mask indicating which coordinates of $W$ are selected \\
$f_z$ 
& Subset predictor fitted on pooled training data \\
$R_e(z)$ 
& Predictive risk of subset $z$ in environment $e$ \\
$z^\star(e)$ 
& Optimal covariate subset in environment $e$\\
$u_e$ 
& Environment representation in environment $e$ \\
$\widehat R_e(z)$ 
& Empirical risk of subset $z$ in training environment $e$ \\
$\widehat g$ 
& Learned gate mapping $u_e$ to weights over $Z$ \\
$f_{\mathrm{env}}$ 
& Encoder mapping the covariates of an environment to $u_e$ \\
\bottomrule
\end{tabular}
\caption{Notation for \Cref{sec:aefs} and \Cref{alg:eafs1,alg:eafs2}.}
\label{tab:eacs_notation}
\end{table}

\subsection{Risk-based covariate selection}
\label{subsec:risk_based}

For each environment $e \in \mathcal{E}$, let $\mathbb{P}_e$ denote the joint distribution of the observed variables $(W, Y)$. The goal is to select a subset of covariates that minimizes prediction error under $\mathbb{P}_e$. We work with a finite library $Z \subseteq \{0,1\}^{p}$ of candidate subsets, where each $z \in Z$ is a binary mask over the $p$ observed covariates $(C_s, X)$. For each $z$, a subset predictor $f_z$ is trained on the pooled labeled data from all training environments using the covariates selected by $z$, producing a library $\{f_z : z \in Z\}$.

The environment-specific risk of subset $z$ is
\[
R_e(z) = \mathbb{E}_{e}\!\left[(Y - f_z(W_z))^2\right],
\]
where $W_z$ denotes the coordinates of $W$ selected by $z$. The optimal subset in environment $e$ is
\[
z^\star(e) = \arg\min_{z \in Z} R_e(z),
\]
which would be the best choice if the true risks were known. 

Because outcomes are unavailable at test time, EACS does not estimate $z^\star(e)$ from labels in the target environment. Instead, it learns from the training environments how the subset risks vary with the covariate distribution.

\subsection{Environment representations}
\label{subsec:env_repr}

Each environment is summarized by a representation
\[
u_e = f_{\mathrm{env}}\!\big(\{W_{i,e}\}_{i=1}^{n_e}\big)
\]
computed from the observed covariates in environment $e$. The encoder $f_{\mathrm{env}}$ maps a set of covariate samples to a fixed-dimensional summary of the covariate distribution, and its role is to recover information about the intervention $i(e)$, on which $z^\star(e)$ depends; \Cref{subsec:linear_scm} discusses when simple summaries carry this information.

The encoder $f_{\mathrm{env}}$ can take several forms. In simple problems, a small set of hand-crafted summaries may be sufficient, as in the running example where the risk comparison depends on variances or correlations. In more complex settings, the relevant signal may involve nonlinearities, interactions, higher-order moments, or other distributional features that are not known in advance. EACS can then learn the environment representation from covariate samples, for example using a permutation-invariant encoder such as DeepSets \citep{zaheer2017deep}.

\subsection{Learning the selector from subset risks}
\label{subsec:learn_selector}

The selector is a gate $g$ that maps an environment representation to weights $g(u_e) = (g_z(u_e))_{z \in Z}$ that are nonnegative and sum to one; a one-hot gate selects a single subset. In each training environment, the library is evaluated on all labeled observations, giving the vector of empirical risks $\big(\widehat R_e(z)\big)_{z \in Z}$. The gate is chosen from a class $\mathcal{G}$ by minimizing the \emph{empirical selection risk}
\begin{equation}
\widehat{\mathcal{S}}(g) = \frac{1}{m}\sum_{j=1}^{m} \sum_{z \in Z} g_z(u_{e_j})\, \widehat R_{e_j}(z),
\label{eq:selection-risk}
\end{equation}
where $m$ is the number of training environments. Each environment receives equal weight, and for a one-hot gate $\widehat{\mathcal{S}}(g)$ is the average risk of the selected predictors.

Using the full risk vector, rather than only the best subset in each environment, means that choosing a near-tied subset costs little, whereas choosing a clearly inferior one costs much.

\subsection{Prediction}
\label{subsec:prediction_interp}

At test time, given a new environment $e^* \in \mathcal{E}$, EACS computes its representation $u_{e^*}$ and the weights $\widehat g(u_{e^*})$, which are shared by all samples in the environment. It then predicts with the largest-weight subset predictor (hard selection) or with the weighted average $\sum_{z} \widehat g_z(u_{e^*}) f_z$ (soft averaging).

\subsection{Discrete and continuous implementations}
\label{subsec:implementations}

EACS admits a discrete and a continuous implementation, which differ in both the selector form and the training scheme. The discrete version selects or averages the predictors of a finite library, using a gate trained on their risks; it is interpretable and admits the theoretical analysis of \Cref{subsec:theorem}. The continuous version weights individual covariates within a shared predictor, which is trained jointly with the gate to minimize prediction loss rather than the selection risk \eqref{eq:selection-risk}; it is not covered by these guarantees but scales to high-dimensional settings. Either implementation may use a hand-crafted or a learned encoder.

\noindent\textbf{Discrete selector.}
\Cref{alg:eafs1} summarizes the discrete version. The gate can be any model whose outputs are weights over $Z$, such as a softmax model or a tree ensemble.

\begin{algorithm}[t]
\caption{EACS (discrete selector)}\label{alg:eafs1}
\textbf{Input:} Training environments $\mathcal{E}_{\mathrm{train}} \subseteq \mathcal{E}$ with labeled data $\{(W_{i,e}, Y_{i,e})\}_{i=1}^{n_e}$ for each $e \in \mathcal{E}_{\mathrm{train}}$; candidate library $Z$; environment encoder $f_{\mathrm{env}}$; gate class $\mathcal{G}$; prediction rule (hard or soft).
\begin{algorithmic}[1]
\Statex \textbf{Training:}
\For{each $z \in Z$}
  \State Fit predictor $f_z$ on the pooled training data, using only the covariates selected by $z$, and hold it fixed.
\EndFor
\For{each $e \in \mathcal{E}_{\mathrm{train}}$}
  \State Compute environment representation $u_e \gets f_{\mathrm{env}}\!\big(\{W_{i,e}\}_{i=1}^{n_e}\big)$.
  \State Compute empirical risk for each $z \in Z$:
  \[
        \widehat R_e(z) \gets \frac{1}{n_e} \sum_{i=1}^{n_e} \big(Y_{i,e} - f_z((W_{i,e})_z)\big)^2.
  \]
\EndFor
\State Fit the gate $\widehat g \in \mathcal{G}$ by minimizing the empirical selection risk \eqref{eq:selection-risk}.

\medskip
\Statex \textbf{Test time:} given a new environment $e^* \in \mathcal{E}$ with unlabeled covariates $\{W_{i,e^*}\}_{i=1}^{n_{e^*}}$,
\Statex \quad (a) compute $u_{e^*} = f_{\mathrm{env}}\!\big(\{W_{i,e^*}\}_{i=1}^{n_{e^*}}\big)$ and the weights $\widehat g(u_{e^*})$;
\Statex \quad (b) for each sample $i$, predict with $f_{\widehat z_{e^*}}$, where $\widehat z_{e^*} = \arg\max_{z} \widehat g_z(u_{e^*})$ (hard selection), or with $\sum_{z} \widehat g_z(u_{e^*}) f_z$ (soft averaging).
\end{algorithmic}
\end{algorithm}

The library $Z$ is typically structured and small: for instance, all subsets of size at most $k$, or unions of a fixed causal core with a small number of plausible proxy variables. This keeps both the computational cost and the complexity of the selection problem manageable. Enumeration becomes prohibitive as the number of covariates grows, motivating the continuous implementation below.

\noindent\textbf{Continuous feature gating.}
To avoid enumerating $Z$, the continuous implementation (\Cref{alg:eafs2}) replaces the binary mask with a vector of feature weights $\tilde z_e \in (0,1)^p$ that depends on $u_e$. Predictions are formed by element-wise weighting of the covariates in a shared predictor, and the predictor, the gating network, and any learned encoder are trained jointly to minimize the environment-averaged prediction loss.

\begin{algorithm}[t]
\caption{EACS via continuous feature gating}\label{alg:eafs2}
\textbf{Input:} Training environments $\mathcal{E}_{\mathrm{train}} \subseteq \mathcal{E}$ with labeled data $\{(W_{i,e}, Y_{i,e})\}_{i=1}^{n_e}$ for each $e \in \mathcal{E}_{\mathrm{train}}$; shared predictor $p_{\theta_p}$; gating network $f_{\mathrm{gate}}$ with parameters $\theta_{\mathrm{gate}}$; environment encoder $f_{\mathrm{env}}$ with parameters $\theta_{\mathrm{env}}$ (empty for a fixed summary); temperature $\tau > 0$.
\begin{algorithmic}[1]
\Statex \textbf{Training:}
\State Initialize $\vartheta = (\theta_p, \theta_{\mathrm{gate}}, \theta_{\mathrm{env}})$.
\State For each $e \in \mathcal{E}_{\mathrm{train}}$, define the environment-specific feature weights
\[
\tilde z_e \;=\; \sigma\!\left(\frac{f_{\mathrm{gate}}\!\big(f_{\mathrm{env}}(\{W_{i,e}\}_{i=1}^{n_e}; \theta_{\mathrm{env}}); \theta_{\mathrm{gate}}\big)}{\tau}\right) \;\in\; (0,1)^p,
\]
where the sigmoid function $\sigma$ is applied element-wise and $\tilde z_e$ is shared across all samples in environment $e$.
\State Minimize the environment-averaged loss
\[
\mathcal{L}(\vartheta) \;=\; \frac{1}{|\mathcal{E}_{\mathrm{train}}|} \sum_{e \in \mathcal{E}_{\mathrm{train}}} \frac{1}{n_e} \sum_{i=1}^{n_e} \mathcal{L}_{\mathrm{pred}}\!\big(Y_{i,e},\, p_{\theta_p}(\tilde z_e \circ W_{i,e})\big)
\]
jointly over $\vartheta$ by gradient-based optimization, adding a regularization penalty if desired.

\medskip
\Statex \textbf{Test time:} given a new environment $e^* \in \mathcal{E}$ with unlabeled covariates $\{W_{i,e^*}\}_{i=1}^{n_{e^*}}$,
\Statex \quad (a) compute $u_{e^*} = f_{\mathrm{env}}\!\big(\{W_{i,e^*}\}_{i=1}^{n_{e^*}}; \hat\theta_{\mathrm{env}}\big)$;
\Statex \quad (b) form $\tilde z_{e^*} = \sigma\!\big(f_{\mathrm{gate}}(u_{e^*}; \hat{\theta}_{\mathrm{gate}})/\tau\big)$;
\Statex \quad (c) predict $\widehat Y_{i,e^*} = p_{\hat\theta_p}(\tilde z_{e^*} \circ W_{i,e^*})$ for each sample $i$.
\end{algorithmic}
\end{algorithm}

The weights $\tilde z_e$ depend only on $u_e$ and are shared across all samples in environment $e$, so covariate selection remains an environment-level operation. As the temperature $\tau$ decreases, the sigmoid becomes increasingly step-like, but training need not produce binary masks, and we do not interpret the fitted weights as recovered causal supports.

\subsection{Incorporating prior causal knowledge}
\label{sec:causal-selector}

When some covariates are known to be causal parents of $Y$, requiring them to appear in every candidate subset simplifies the selection problem. Let $S \subseteq \{1, \dots, p\}$ denote the indices of these covariates. We impose this constraint by restricting the library to
\[
Z_S \;=\; \{ z \in Z : z_j = 1 \text{ for all } j \in S \}.
\]
If $Z$ contains all subsets, $|Z_S| = 2^{p - |S|}$. In the running example, requiring $C_2$ reduces the library from the four subsets of $\{C_2, X\}$ to $\{C_2\}$ and $\{C_2, X\}$.

The guarantee of \Cref{subsec:theorem} applies to the constrained selector with the library and gate class restricted accordingly. Because the restricted library and gate class are smaller, its error terms can only decrease, but the best gate in the restricted class can be worse: the constrained selector cannot drop a covariate in $S$, even in environments where it hurts prediction. This can happen if the covariate is not in fact a causal parent, but also if it is a causal parent that is correlated with an unobserved parent, since its pooled coefficient then partly reflects that parent.

\paragraph{Constrained discrete and continuous selectors.}
The constrained variants of \Cref{alg:eafs1,alg:eafs2} are simple modifications: the discrete selector assigns weights only to subsets in $Z_S$, and the continuous version fixes the feature weights for $j \in S$ at one while learning environment-dependent weights for the remaining covariates. In our simulations, the constraint lowers prediction error when outcome noise is high and either the training environments or the observations per environment are few; otherwise, the constrained and unconstrained selectors perform similarly. Full algorithms, theoretical statements, and simulation results are given in \Cref{sec:s_causal_full}.

\section{When Does EACS Work?}
\label{sec:analysis}  

\Cref{subsec:theorem} gives theoretical guarantees, \Cref{rem:env_specific} explains why EACS restricts adaptation to the choice of covariates, \Cref{subsec:linear_scm} discusses when environment summaries carry the information needed for selection, and \Cref{subsec:sim33} returns to the running example.  

\subsection{Theoretical guarantees}
\label{subsec:theorem}
For the discrete selector of \Cref{alg:eafs1}, we bound the risk of the learned gate on a new environment relative to the best gate in the searched class. To do so, we assume that the training environments $e_1, \dots, e_m$ and the test environment $e^*$ are drawn i.i.d.\ from a distribution $\mathcal{D}$ over the shift class $\mathcal{E}$. For simplicity, each training environment contains $n$ i.i.d.\ labeled observations, and the test representation $u_{e^*}$ is computed from $n$ unlabeled observations.

The quantity of interest is the selection risk of a gate $g$ on a new environment,
\[
\mathcal{S}(g) = \mathbb{E}_{e^* \sim \mathcal{D}}\Big[\sum_{z \in Z} g_z(u_{e^*})\, R_{e^*}(z)\Big],
\]
the population version of the empirical selection risk \eqref{eq:selection-risk}, where the expectation is also over the unlabeled batch that determines $u_{e^*}$. For a one-hot gate, $\mathcal{S}(g)$ is the risk of the selected predictors; for soft averaging, the risk is at most $\mathcal{S}(g)$ by convexity of the squared loss.

Let $\mathcal{F}$ be a fixed hypothesis class that contains every library, encoder, and gate the procedure can fit.

\begin{theorem}[Finite-sample bound]\label{theorem:finite}
Suppose that the environments are drawn i.i.d.\ from $\mathcal{D}$ and that every squared loss $(Y - f_z(W_z))^2$ of a library in $\mathcal{F}$ lies in $[0, B]$. Let $\widehat{\mathcal{G}}$ be the class of gates searched by the procedure, and let $\widehat g \in \widehat{\mathcal{G}}$ satisfy $\widehat{\mathcal{S}}(\widehat g) \le \inf_{g \in \widehat{\mathcal{G}}} \widehat{\mathcal{S}}(g) + \varepsilon_{\mathrm{opt}}$. For any $\delta \in (0,1)$, with probability at least $1 - \delta$ over the training data,
\[
\mathcal{S}(\widehat g) - \inf_{g \in \widehat{\mathcal{G}}} \mathcal{S}(g)
\;\le\; 4 r_m + 2B\sqrt{\frac{\log(1/\delta)}{2m}} + 2 a_n + \varepsilon_{\mathrm{opt}},
\]
where $r_m$ and $a_n$ are defined below.
\end{theorem}

The bound has three sources of error. The first two terms reflect learning the gate from $m$ environments, where $r_m$ is the Rademacher complexity over $\mathcal{F}$ of the environment-level losses $\sum_z g_z(u_e)\, \widehat R_e(z)$. The third term reflects estimating each environment's risks from $n$ observations, with $a_n = \mathbb{E} \sup_{\mathcal{F}} \max_z |R_e(z) - \widehat R_e(z)|$. The last term is the optimization gap, which is zero when the gate is found by exhaustive search over a finite class. Because the bound is uniform over $\mathcal{F}$, it holds even when the library, encoder, and gate are fitted on the same data. Since the summaries are computed from finite samples, it also accounts for their sampling noise, which adds no error term but can make the best gate in the class worse, because noisy summaries separate environments less well. The boundedness assumption can be relaxed to moment conditions by a standard truncation argument. Proofs are in \Cref{sec:s_proofs}.

For classes of controlled capacity, $r_m$ vanishes as $m$ grows and $a_n$ as $n$ grows.

\begin{cor}[Consistency]\label{theorem:asymptotic}
Under the conditions of \Cref{theorem:finite}, suppose that $m, n \to \infty$, $r_m \to 0$, $a_n \to 0$, and $\varepsilon_{\mathrm{opt}} \to 0$ in probability. Then
\[
\mathcal{S}(\widehat g) - \inf_{g \in \widehat{\mathcal{G}}} \mathcal{S}(g) \;\xrightarrow{\mathbb{P}}\; 0.
\]
\end{cor}

\paragraph{What the guarantee requires.}
Assuming i.i.d.\ environments is the standard way to state a guarantee for new environments \citep{blanchard2021domain}, and the number of environments needed then depends on what is learned across them. For EACS, the problem learned across environments is simple: the gate only chooses among a few predictors fitted on all pooled observations, so $r_m$ is small when the library and gate class are small. The training environments must show when proxies help and when they hurt, but they need not cover every shift encountered at test time (\Cref{subsec:sim33,sec:contextual_main}). This learning step does not use the causal model, so the guarantee holds without knowing the causal graph or which covariates are causal. The causal structure instead determines what is learned and whether it can succeed: it motivates the library of subset predictors, determines whether the summaries carry the information needed to choose among them (\Cref{subsec:linear_scm}), and supplies constraints when causal parents are known (\Cref{sec:causal-selector}).

\subsection{Why restrict adaptation to the choice of covariates?} 
\label{rem:env_specific}

A contextual predictor appends $u_e$ to each observation and learns $h(W_{i,e}, u_e)$, so that every coefficient can vary with the environment. For instance, the linear contextual predictor \[a + \beta^\top W + \gamma^\top u_e + W^\top B u_e\] makes each coefficient an affine function of the summary. This flexibility helps when covariate effects change across environments: the best weight on a proxy, for example, decreases as the proxy becomes noisier, and the contextual predictor can track this change, whereas EACS can only select or average fixed predictors, whose coefficients are compromises across all training environments. 

The flexibility also has two costs, which EACS avoids. First, the contextual predictor adapts every coefficient, including those of covariates whose effects do not change, and each adaptation must be estimated from the training environments; EACS fits these coefficients once on the pooled observations. Second, the contextual predictor learns how its coefficients depend on the summary only over the range of the training environments, so it must extrapolate to test environments with stronger shifts; EACS always predicts with its pooled predictors or an average of them, so even for an unfamiliar summary it behaves at worst like one of these fixed predictors. In the continuous version, the feature weights lie in $(0,1)$, so they can only shrink the coefficients of a linear shared predictor. 

EACS is therefore expected to perform better when many covariates need no adaptation or when the training shifts do not cover the test shifts, and the contextual predictor otherwise. Restricting adaptation also keeps the selection interpretable and allows causal constraints on $Z$. \Cref{sec:contextual_main} tests these predictions in a controlled simulation, and \Cref{sub:Bike,sub:ACS} compare both approaches on real data.

\subsection{When are environment summaries informative?}
\label{subsec:linear_scm}

\Cref{theorem:finite} compares $\widehat g$ with the best gate in the searched class. For this gate to choose well in each environment, the summary must reflect the shifts that change which subset is best. For linear library predictors $f_z(W_z) = a_z + \beta_z^\top W$ as in the running example, where $\beta_z$ is zero outside the coordinates selected by $z$, expanding the square gives
\[
R_e(z) = \Var_e(Y) + \beta_z^\top \Sigma_{W,e} \beta_z - 2\beta_z^\top \Sigma_{WY,e} + \big(\mu_{Y,e} - a_z - \beta_z^\top \mu_{W,e}\big)^2,
\]
where $\mu_{W,e}$ and $\Sigma_{W,e}$ are the mean and covariance of $W$, $\mu_{Y,e} = \mathbb{E}_e(Y)$, and $\Sigma_{WY,e} = \Cov_e(W, Y)$. The first term is common to all subsets, so the comparison between subsets depends on two kinds of moments: covariate moments, which can be estimated from unlabeled target data, and outcome moments $\mu_{Y,e}$ and $\Sigma_{WY,e}$, which cannot. A summary of the covariate moments is therefore informative when environments with the same covariate moments also have the same outcome moments. For nonlinear predictors, the risks depend on more than the first two moments, so the summary may need to capture more of the covariate distribution, but the same reasoning applies.

\paragraph{Linear structural causal models.}
The causal structure determines when this holds. Consider a linear structural causal model in which each environment shifts the means of some covariates, observed or unobserved, or adds independent noise to them, leaving the equation for $Y$ unchanged. Each such intervention leaves a \emph{trace} on the observed covariates: a mean shift changes their means, and added noise changes their covariance. The affected covariates are the intervened variable itself, if it is observed, and its observed descendants, each changed by an amount determined by the total causal effect of the intervened variable on that covariate.

\begin{prop}\label{prop:linear_scm}
In such a model, suppose that
\begin{enumerate}[label=(\roman*), noitemsep, topsep=2pt]
\item whenever two ways of shifting means change the mean of $W$ in the same way, they also change the mean of $Y$ in the same way; and
\item whenever two ways of adding noise change the covariance of $W$ in the same way, they also change $\Cov_e(W, Y)$ in the same way.
\end{enumerate}
Then, for every linear library predictor, $R_e(z) - \Var_e(Y)$ depends on the environment only through $\mu_{W,e}$ and $\Sigma_{W,e}$. Hence, the covariate mean and covariance determine which subset is best in each environment.
\end{prop}

In words, interventions that leave the same trace must have the same effect on the outcome moments. Since an intervention on an observed covariate changes that covariate itself, the conditions can fail only through an unobserved parent, which is visible only through its observed descendants, that is, through its proxies. They fail when an intervention on the parent and one on its proxies leave the same trace but affect the outcome moments differently. If the two interventions also favor different subsets, no summary of the covariate moments can choose correctly under both, as the running example below shows.

\begin{figure}[t]
\centering
\definecolor{EACSBlue}{HTML}{0072B2}
\definecolor{EACSInk}{HTML}{333333}
\definecolor{EACSHidden}{HTML}{E6E6E6}
\definecolor{EACSOrange}{HTML}{D55E00}
\begin{tikzpicture}[
    var/.style={circle, draw=EACSInk, line width=0.55pt, fill=white,
      minimum size=7.2mm, inner sep=0pt},
    hidden/.style={var, fill=EACSHidden},
    edge/.style={-{Stealth[length=1.7mm, width=1.15mm]}, draw=EACSInk, line width=0.65pt},
    perturb/.style={-{Stealth[length=2mm, width=1.4mm]}, draw=EACSBlue, line width=1.1pt},
    trace/.style={draw=EACSOrange, line width=1.1pt, dashed},
    title/.style={font=\sffamily\small, anchor=base},
    lab/.style={font=\sffamily\footnotesize, anchor=base}
]
\foreach \dx/\t/\heading/\effect in {
    0/a/{Mean shift on $C_1$}/{Mean of $X$ changes},
    4.9/b/{Noise on $X$}/{Variance of $X$ increases},
    9.8/c/{Noise on $C_1$}/{Variance of $X$ increases}} {
  \begin{scope}[xshift=\dx cm]
    \node[title] at (1.4, 1.82) {(\t) \heading};
    \node[hidden] (C1\t) at (1.4, 0.78)  {$C_1$};
    \node[var]    (C2\t) at (1.4, -0.78) {$C_2$};
    \node[var]    (X\t)  at (0, 0)       {$X$};
    \node[var]    (Y\t)  at (2.8, 0)     {$Y$};
    \foreach \a/\b in {C1/X, C1/Y, C2/X, C2/Y}
      \draw[edge] (\a\t) -- (\b\t);
    \node[lab] at (1.4, -1.75) {\effect};
  \end{scope}
}
\draw[EACSBlue, line width=1.1pt] (C1a) circle (3.6mm);
\draw[perturb] ([yshift=4mm]C1a.north) -- (C1a.north);
\draw[trace] (Xa) circle (4.4mm);
\draw[EACSBlue, line width=1.1pt] (Xb) circle (3.6mm);
\draw[perturb] ([yshift=4mm]Xb.north) -- (Xb.north);
\draw[trace] (Xb) circle (4.4mm);
\draw[EACSBlue, line width=1.1pt] (C1c) circle (3.6mm);
\draw[perturb] ([yshift=4mm]C1c.north) -- (C1c.north);
\draw[trace] (Xc) circle (4.4mm);
\end{tikzpicture}
\caption{Traces of three shifts in the running example. Blue outlines and arrows mark the intervened variable, and orange dashed outlines mark the observed covariate whose distribution changes. A mean shift on $C_1$ (a) and noise on $X$ (b) leave different traces, so a covariate summary can tell them apart. Noise of the same variance on $C_1$ (c) makes $X$ more, rather than less, informative about $Y$ but leaves the same trace as noise on $X$ (b), so no covariate summary can tell (b) and (c) apart.\label{fig:traces}}
\end{figure}

\paragraph{The running example.}
Here $W = (C_2, X)$ and $U = C_1$. Mean shifts act only on $C_1$, whose trace is the change in the mean of $X$ (\Cref{fig:traces}a). Since the mean of $X$ is never shifted directly, different mean shifts leave different traces, and condition (i) holds. Noise acts on $C_2$ or $X$, whose traces on the covariance of $(C_2, X)$ are distinct, so condition (ii) holds and moment summaries are informative. 

Now suppose that noise could also be added to $C_1$. Noise of variance $w$ on $C_1$ (environment $e_A$) and on $X$ (environment $e_B$) both increase the variance of $X$ by $w$ and leave the other covariate moments unchanged (\Cref{fig:traces}b,c). In this Gaussian model, the two covariate distributions are then identical, but the outcome moments differ, so condition (ii) fails. In $e_A$, the extra variation in $X$ comes from $C_1$, which also drives $Y$, so $X$ becomes more informative about $Y$ and keeping it remains beneficial. In $e_B$, the extra variation is unrelated to $Y$, so $X$ becomes noisier, and for large $w$ dropping it is better (\Cref{sec:sim_analytic_rule}). Since the two environments have the same covariate distribution, no rule based on unlabeled target covariates can choose correctly in both. 

The same problem arises for mean shifts: shifting the mean of $C_1$ or of $X$ by $\delta$ changes the mean of $X$ by $\delta$ and leaves the other covariate moments unchanged, but only the shift on $C_1$ changes the mean of $Y$, so condition (i) fails, and for large $\delta$ the two shifts again favor different subsets. Our main simulations include neither pair of shifts; \Cref{sec:s_ambiguity} simulates each pair and confirms that EACS cannot match the better fixed predictor under both shifts. This limitation is not specific to selecting covariates: the contextual predictor of \Cref{rem:env_specific} also depends on the environment only through the covariate summary, so it cannot be optimal under both shifts either.

\subsection{Running example continued: empirical evaluation of EACS}
\label{subsec:sim33}

We now evaluate EACS in the running example, where the shifts move the mean of $C_1$ or add noise to $C_2$ or $X$. The library contains the four subsets of $\{C_2, X\}$, each fitted by pooled least squares, and the summary $u_e$ consists of the correlation between $C_2$ and $X$ and their SDs. The gate is a decision tree that selects one subset for each environment and is fitted to minimize the selection risk \eqref{eq:selection-risk}; \Cref{sec:s_sim_additional} gives the details.

We vary four factors: the number of training environments, the number of observations per environment, the informativeness of $u_e$, and the perturbation coverage during training. EACS keeps the proxy where it helps and drops it where it hurts, and it achieves lower error than both fixed predictors (\Cref{fig:sim_a_env_mse} and \Cref{tab:sim_summary} in \Cref{sec:s_sim_additional}). This holds even with few training environments when outcome noise is low, and even when the training shifts are weaker than those at test time, because EACS only needs to learn when the proxy becomes unreliable. When environments or observations are few and outcome noise is high, however, it can be worse than always keeping the proxy (\Cref{fig:sim_a_env_mse,fig:sim_a_sam_mse}). EACS cannot learn when to drop the proxy if no training environment disrupts it or if the summary omits the informative statistics.

\section{Empirical Studies}
\label{sec:applications}

We compare EACS with two kinds of methods. Static baselines use a single predictor for all environments: least squares, lasso, anchor regression, and, in the bike-sharing study, invariant causal prediction (ICP). They show whether adapting to the environment helps. The linear contextual predictor of \Cref{rem:env_specific} also adapts, using the same summary $u_e$ as EACS, or an encoder of the same architecture when the summary is learned. For a fixed summary, both this predictor and EACS are linear in the covariates, so comparing them shows whether restricting adaptation to the choice of covariates is worthwhile.

We first compare EACS with the contextual predictor in a controlled simulation, and then compare it with both kinds of methods in two empirical studies under complementary regimes. In the bike-sharing study, each calendar day is an environment, giving many environments with few observations each; this study uses the discrete selector of \Cref{alg:eafs1}. In the 2018 American Community Survey (ACS) Income study, each state is an environment, giving fewer environments with many observations and many more covariates; this study uses the continuous version of \Cref{alg:eafs2}. Implementation details for both studies are given in \Cref{sec:implementation}. 

\subsection{Subset selection versus contextual prediction: a controlled comparison}
\label{sec:contextual_main}

We compare the two approaches in the running example of \Cref{section:running-example-init}, with the same shifts and test environments as in \Cref{subsec:sim33} but independent replications. In some settings, we add irrelevant covariates: independent standard normals that are unrelated to the outcome and unaffected by the shifts. They are the simplest example of covariates that need no adaptation across environments, like many candidate features collected in practice. Starting from a reference setting with $20$ training environments per shift type, $100$ observations per environment, no irrelevant covariates, training outcome-noise SD $1$, and training perturbations up to $4$, we vary one factor at a time: the number of observations per environment, the number of training environments, the number of irrelevant covariates, the training outcome noise, the content of the summary, and the range of training perturbations. We focus here on the two factors that correspond to the two costs of contextual prediction in \Cref{rem:env_specific}: the number of irrelevant covariates, which need no adaptation, and the range of training perturbations; \Cref{sec:s_contextual_all} reports all six.

Both methods use the same summary: the SDs and pairwise partial correlations of all observed covariates, including the irrelevant ones. EACS is fitted as in \Cref{subsec:sim33}, with a library of all subsets of the observed covariates; the contextual predictor is fitted by ridge regression, with the penalty chosen by cross-validation over training environments.

\begin{figure}[t]
\centering
\includegraphics[width=\linewidth]{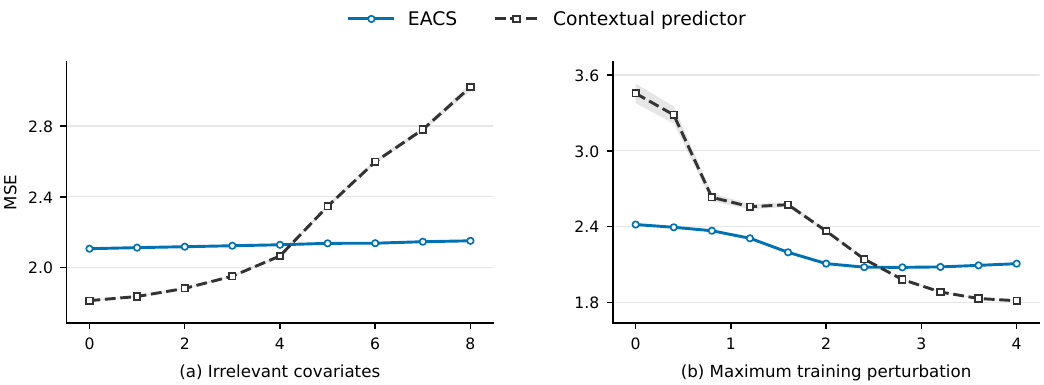}
\caption{EACS versus the contextual predictor in the running example as (a) the number of irrelevant covariates and (b) the largest training perturbation vary from the reference setting. Points show mean test MSE over 1{,}000 replications, with pointwise 95\% confidence intervals.}
\label{fig:contextual_linear_comparison}
\end{figure}

\begin{table}[t]
\centering
\caption{Test MSE of EACS and the contextual predictor as the number of irrelevant covariates and the range of training perturbations vary (\Cref{fig:contextual_linear_comparison}). The reduction is $100(1 - \text{EACS}/\text{contextual})$; negative values favor the contextual predictor.}
\label{tab:contextual_summary}
\begin{tabular}{lccc}
\toprule
Setting & EACS & Contextual & Reduction (\%) \\
\midrule
Reference & 2.11 & 1.81 & $-16.2$ \\
\addlinespace
4 irrelevant covariates & 2.13 & 2.07 & $-3.1$ \\
5 irrelevant covariates & 2.14 & 2.35 & 8.9 \\
8 irrelevant covariates & 2.15 & 3.02 & 28.8 \\
\addlinespace
Training perturbations up to 2.4 & 2.08 & 2.14 & 3.0 \\
Training perturbations up to 2 & 2.11 & 2.36 & 10.9 \\
Training perturbations up to 0.8 & 2.37 & 2.63 & 10.1 \\
No training perturbations & 2.42 & 3.45 & 30.1 \\
\bottomrule
\end{tabular}
\end{table}

\Cref{fig:contextual_linear_comparison} and \Cref{tab:contextual_summary} show that neither method is uniformly better. In the reference setting, the best coefficients change with the shift and the training shifts cover the test shifts, so the flexibility of the contextual predictor pays off: it is more accurate than EACS, and it remains so when we vary the number of observations per environment, the number of training environments, the training noise, or the summary (\Cref{sec:s_contextual_all}). The two costs identified in \Cref{rem:env_specific} reverse this ordering. First, irrelevant covariates barely affect EACS but degrade the contextual predictor, which must still learn how their coefficients, and those of the summary components they add, vary across environments; EACS is more accurate once there are five or more of them. Second, when the training shifts are weaker than those at test time, the contextual predictor must extrapolate its coefficients, whereas EACS behaves at worst like one of its fixed predictors; EACS is more accurate when the largest training perturbation is at most $2.4$.

\subsection{Bike-sharing dataset}
\label{sub:Bike}

We evaluate EACS and several baselines on the bike-sharing dataset \citep{fanaee2013bike, fanaee2013event}, which contains $17{,}379$ hourly records of bike rentals in Washington, D.C., during 2011--2012. Following \citet{Rothenhausler2021anchor}, we fit linear models for a transformed rental count using four weather covariates: temperature, feeling temperature, humidity, and wind speed. Each calendar day is treated as an environment; we use the $730$ days with at least three hourly records, which excludes one day with a single record.

The $730$ days are divided into five consecutive blocks of $146$ days. In each outer fold, models are trained on four blocks and evaluated on the held-out block, and we report the mean MSE over all held-out days. Within each outer fold, tuning parameters are selected by three-fold cross-validation (CV) over the training days, and the selected configuration is refitted on all training days. We compare the following methods.
\begin{enumerate}[label=(\alph*), leftmargin=*]
\item \textit{OLS.} Least squares on all four weather covariates; this is also the full-covariate predictor in the EACS library.

\item \textit{Lasso, anchor regression, and ICP.}
The penalty of the lasso \citep{tibshirani1996regression} and the robustness parameter of anchor regression \citep{Rothenhausler2021anchor}, with day indicators as anchors, are tuned by CV. ICP \citep{peters2016causal} uses significance level $0.05$ and predicts from the intersection of the subsets accepted as invariant across days, or from the intercept-only model if no subset is accepted.

\item \textit{EACS.}
The library contains the $16$ subsets of the four weather covariates, including the intercept-only model, each fitted by least squares on the training days. The summary of each day consists of the sample means, sample SDs, and pairwise partial correlations of the four covariates. CV selects the gate among linear softmax models, small neural networks, and random forests, all trained to minimize the selection risk \eqref{eq:selection-risk}, and chooses between hard selection and soft averaging.

\item \textit{Contextual predictor.}
Ridge regression on the covariates, the summary used by EACS, and their interactions, as in \Cref{rem:env_specific}, with the penalty tuned by CV.
\end{enumerate}

\begin{figure}[t]
  \centering
  \includegraphics[width=1\linewidth]{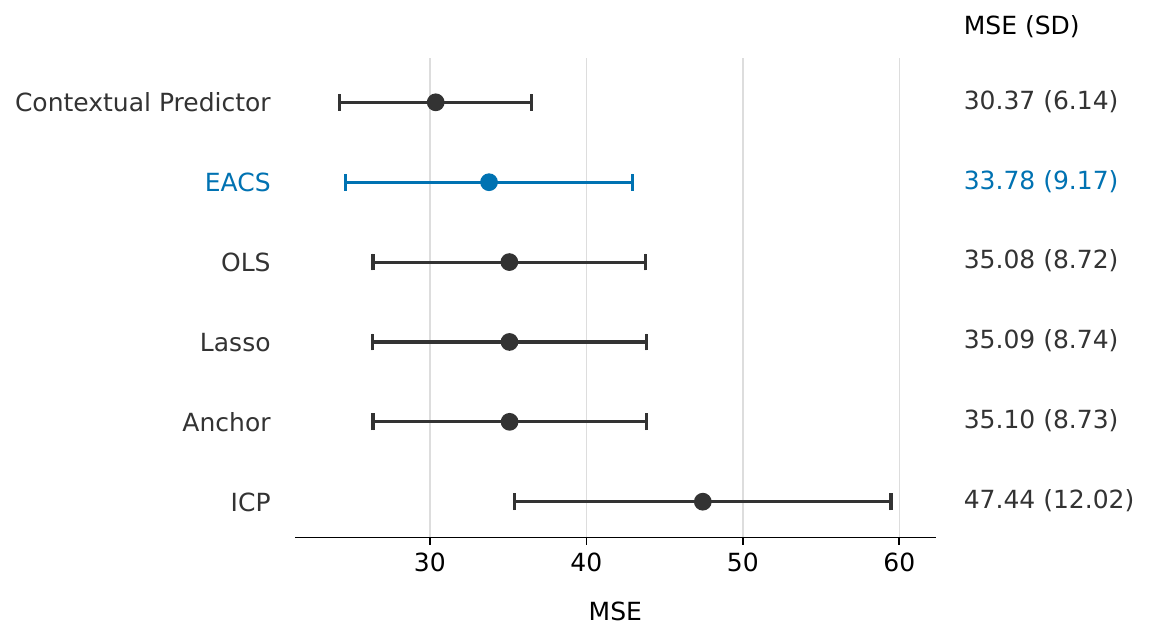}
  \caption{Bike-sharing comparison. Points show the mean MSE over all held-out days, error bars indicate $\pm 1$ SD of the five fold means, and the numbers report the mean (SD).}
  \label{fig:bike_res}
\end{figure}

\noindent\textbf{Results.}
\Cref{fig:bike_res} reports the results. EACS attains a mean MSE of $33.78$ and improves on OLS, lasso, and anchor regression ($35.08$, $35.09$, and $35.10$) and on ICP ($47.44$). Anchor regression selects $\gamma = 1$ in every fold and lasso selects a small penalty, so both nearly coincide with OLS; ICP returns the empty set in every fold and reduces to the intercept-only model. The contextual predictor attains the lowest mean MSE, $30.37$. With only four covariates and training days from both years, it has few coefficients to adapt and little need to extrapolate, and adjusting the coefficients to each day gives it an advantage over selecting or averaging pooled predictors.

\subsection{ACS Income dataset}
\label{sub:ACS}

We evaluate EACS on the 2018 ACS Income data \citep{ding2021retiring}, following the preprocessing pipeline of \citet{jeong2025out}. The data are census records from all US states and Puerto Rico, and the task is to predict individual log income from age, class of worker, education, marital status, occupation, relationship, weekly working hours, sex, and race. Categorical variables are one-hot encoded, giving $590$ covariate columns. Each state is treated as an environment, giving $51$ environments with about $32{,}600$ observations on average.

We divide the $51$ environments, in alphabetical order of their postal abbreviations, into five consecutive blocks of $10$ or $11$ states. In each outer fold, models are trained on four blocks and evaluated on the held-out block, and we report the mean MSE over all held-out states. Tuning parameters are selected by three-fold CV within the training environments, and the selected configuration is refitted on all training states. We compare the following methods.

\begin{enumerate}[label=(\alph*), noitemsep, topsep=0pt, leftmargin=*]
  \item \textit{OLS.}
  Least squares on all covariates.

  \item \textit{Lasso and anchor regression.}
  The lasso penalty and the anchor robustness parameter (state indicators as anchors) are tuned by CV.

  \item \textit{EACS.}
  Because the covariates are high-dimensional, we use the continuous version of \Cref{alg:eafs2}. They also mix numeric and categorical variables, some with many levels, so a compact hand-crafted summary is difficult to choose; the summary is instead learned by a DeepSets encoder \citep{zaheer2017deep} trained jointly with the gate and the predictor.

  \item \textit{Contextual predictor.}
  The same linear form in the covariates, the summary, and their interactions, with the summary learned by an encoder of the same architecture as in EACS and trained jointly with the coefficients.
\end{enumerate}
We omit ICP because exhaustive search over subsets is computationally infeasible in this setting.

\begin{figure}[t]
  \centering
  \includegraphics[width=1\linewidth]{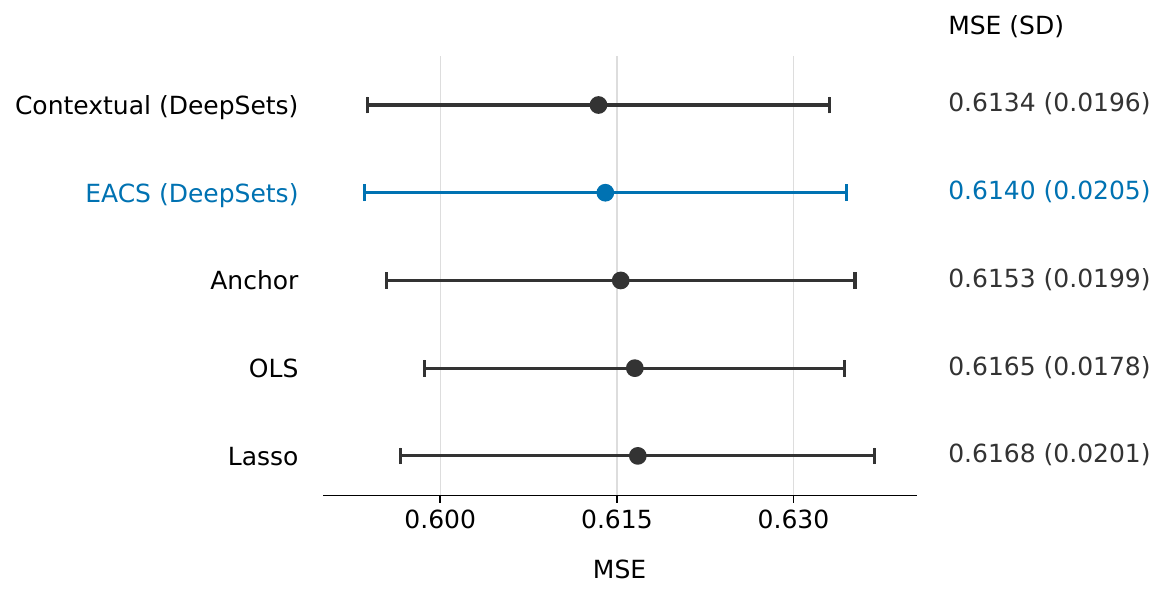}
    \caption{ACS Income comparison. Points show the mean MSE over all held-out states, error bars indicate $\pm 1$ SD of the five fold means, and the numbers report the mean (SD).}
  \label{fig:income_res}
\end{figure}

\noindent\textbf{Results.}
\Cref{fig:income_res} reports the results. EACS attains a mean MSE of $0.6140$ and the contextual predictor $0.6134$, compared with $0.6153$ for anchor regression, $0.6165$ for OLS, and $0.6168$ for lasso. Both adaptive methods improve slightly on the static baselines and are essentially tied, but all these differences are small relative to the variability across folds. The summary matters more for the contextual predictor. With a fixed summary of marginal moments in place of the learned one, EACS remains about as accurate as the best static baseline ($0.6152$), whereas the contextual predictor becomes far less accurate ($1.1792$; \Cref{tab:acs_summary} in \Cref{sec:implementation}), mainly because of Alaska (MSE $24.11$, compared with $0.89$ for EACS and $0.78$ for OLS), whose summary lies far outside the range of the training states. This is the extrapolation cost described in \Cref{rem:env_specific}.

\section{Discussion}

We introduced EACS, an algorithm for environment-adaptive covariate selection under distribution shift. When some causal parents of the outcome are unobserved, non-causal covariates can act as proxies whose usefulness depends on the shift, so no fixed covariate set is uniformly best. EACS learns when to retain or exclude proxy covariates by mapping summaries of the target covariate distribution to weights over covariate-subset predictors, or to feature weights in its continuous version. This is a restricted form of adaptation: the use of covariates varies by environment, while prediction parameters are shared across environments, which keeps the selection interpretable and allows causal constraints. Compared with a contextual predictor, which lets every coefficient vary with the environment, this restriction gives up the ability to track covariate effects that change across environments, but it avoids adapting effects that do not change and extrapolating coefficients beyond the training shifts. In our simulations, EACS therefore performs better when many covariates need no adaptation or when the training shifts do not cover the test shifts, and the contextual predictor performs better otherwise.

The causal analysis clarifies when such adaptation can work. Covariate summaries are informative when the shifts that change which subset is best also leave distinct traces in the observed covariates. When a shift on an unobserved parent leaves the same trace as a shift on its proxy but favors a different subset, no rule based on unlabeled target covariates can choose correctly in both cases, and contextual prediction faces the same limit. Resolving such ambiguities requires information beyond the covariates, such as domain knowledge about which interventions may occur.

The approach has several limitations. Its success depends on informative environment summaries and on training environments in which some shifts disrupt the proxy relationship; without such environments, EACS cannot learn when to drop the proxy, although the training shifts need not span the full range of shifts encountered at test time. The continuous version scales to high-dimensional settings, but it uses a different predictor class and joint optimization, so the discrete selector's guarantees do not apply. EACS also requires a batch of test covariates to construct environment-level summaries, which precludes single-observation prediction, and very small batches give noisy summaries. Finally, misspecified causal constraints can harm performance by forcing incorrect covariates to remain active. Two natural directions for future work are to incorporate uncertain causal knowledge and to combine subset selection with contextual prediction, for example by adding a contextual predictor to the library.

\section*{Data Availability Statement}
Data and code to reproduce all analyses are available at
\href{https://github.com/sushi133/Environment-Adaptive-Covariate-Selection}{\nolinkurl{https://github.com/sushi133/Environment-Adaptive-Covariate-Selection}}.

\section*{Acknowledgments} This work was supported in part by funding from the Office of Naval Research under grant N00014-23-1-2590; the National Science Foundation under grants No.~2310831, No.~2428059, No.~2435696, and No.~2440954; a Michigan Institute for Data Science Propelling Original Data Science (PODS) grant; LG Management Development Institute AI Research; and Two Sigma Investments LP\@. Any opinions, findings, and conclusions or recommendations expressed in this material are those of the authors and do not necessarily reflect the views of the sponsors. The authors declare no competing interests.

\clearpage
\appendix

\section{Derivation of the Analytic Selection Rule}
\label{sec:sim_analytic_rule}

This section derives the analytic selection rule stated in \Cref{section:running-example-init}. As there, we start from the linear Gaussian model
\[
Y = C_1 + C_2 + \varepsilon_Y,
\qquad
X = C_1 - C_2 + \varepsilon_X,
\]
with independent $C_1, C_2, \varepsilon_X \sim \mathcal N(0,1)$ and $\varepsilon_Y \sim \mathcal{N}(0, \sigma_Y^2)$; \Cref{section:running-example-init} uses $\sigma_Y = 1$, and $\sigma_Y^2$ cancels from the selection rule. Environments are generated by perturbing one variable at a time: shifting the mean of $C_1$ by $\delta$, or adding independent Gaussian noise $\mathcal{N}(0,\delta^2)$ to $C_2$ or to $X$, with $\delta \in [0,4]$. These perturbations keep $C_1$, $C_2$, $\varepsilon_X$, and $\varepsilon_Y$ independent, with $\mathrm{Var}_e(C_1) = 1$ and $\mathbb{E}_e(C_2) = \mathbb{E}_e(\varepsilon_X) = 0$ in every environment $e$. Write
\[
\mu_e = \mathbb{E}_e(C_1), \qquad v_{2,e} = \mathrm{Var}_e(C_2), \qquad q_e = \mathrm{Var}_e(\varepsilon_X),
\]
let $s_{2,e}$ and $s_{3,e}$ be the standard deviations (SDs) of $C_2$ and $X$, and let $r_e = \Corr_e(C_2, X)$. Then
\[
s_{2,e}^2 = v_{2,e}, \qquad s_{3,e}^2 = 1 + v_{2,e} + q_e, \qquad
r_e = \frac{\Cov_e(C_2,X)}{s_{2,e}s_{3,e}} = -\frac{s_{2,e}}{s_{3,e}}.
\]

\paragraph{Pooled predictors.}
Both predictors are population least-squares fits with intercepts on the pooled training data. Let $\mu_0$ and $V_1$ be the pooled mean and variance of $C_1$, and $\bar q$ the pooled mean of $q_e$. Since $C_2$ and $\varepsilon_X$ have mean zero in every environment, $C_1$, $C_2$, and $\varepsilon_X$ are uncorrelated in the pooled data, and the pooled means of $Y$ and $X$ both equal $\mu_0$. Hence $\widehat Y_{\{C_2\}} = \mu_0 + C_2$, and for $\{C_2, X\}$ the normal equations give
\[
\widehat Y_{\{C_2,X\}} = (1-\beta_3)\mu_0 + (1+\beta_3) C_2 + \beta_3 X,
\qquad
\beta_3 = \frac{V_1}{V_1 + \bar q} \in (0,1).
\]

\paragraph{Risks.}
Substituting $Y = C_1 + C_2 + \varepsilon_Y$ and $X = C_1 - C_2 + \varepsilon_X$ into the residuals gives
\begin{align*}
Y - \widehat Y_{\{C_2\}} &= (C_1 - \mu_0) + \varepsilon_Y,\\
Y - \widehat Y_{\{C_2,X\}} &= (1-\beta_3)(C_1 - \mu_0) + \varepsilon_Y - \beta_3 \varepsilon_X,
\end{align*}
so
\begin{align*}
R_e(\{C_2\}) &= 1 + (\mu_e - \mu_0)^2 + \sigma_Y^2,\\
R_e(\{C_2,X\}) &= (1-\beta_3)^2\big\{1 + (\mu_e - \mu_0)^2\big\} + \beta_3^2 q_e + \sigma_Y^2.
\end{align*}
The term $\sigma_Y^2$ cancels in the difference. Using $q_e = s_{3,e}^2 - s_{2,e}^2 - 1$,
\begin{align*}
\Delta_e &:= R_e(\{C_2\}) - R_e(\{C_2,X\})\\
&= 2\beta_3 - \beta_3^2\,(s_{3,e}^2 - s_{2,e}^2) + \beta_3(2-\beta_3)(\mu_e - \mu_0)^2,
\end{align*}
which is the expression reported in the main text. Since $\beta_3 > 0$, the $\{C_2\}$ model achieves lower risk exactly when
\[
\Delta_e < 0
\quad\Longleftrightarrow\quad
s_{3,e}^2 - s_{2,e}^2 > \frac{2 + (2-\beta_3)(\mu_e - \mu_0)^2}{\beta_3}.
\]

In mean-shift environments, $s_{3,e}^2 - s_{2,e}^2 = 2 < 2/\beta_3$, so the proxy is always retained there; mean shifts change the size of the risk gap but never reverse it. 
\paragraph{Correlation form.}
Since $r_e^2 = s_{2,e}^2 / s_{3,e}^2$, we have $s_{3,e}^2 - s_{2,e}^2 = s_{3,e}^2 (1 - r_e^2)$, so
\[
\Delta_e < 0
\quad\Longleftrightarrow\quad
1 - r_e^2 > \frac{2 + (2-\beta_3)(\mu_e - \mu_0)^2}{\beta_3\, s_{3,e}^2}.
\]
This is the correlation-based form referenced in the main text.

\paragraph{Noise on the unobserved parent.}
\Cref{subsec:linear_scm} compares two environments with $\mu_e = 0$ and $\mathrm{Var}_e(C_2) = 1$, in which noise of variance $w$ is added to $C_1$ (environment $e_A$) or to $X$ (environment $e_B$). In both, $(C_2, X)$ is Gaussian with mean zero, $\mathrm{Var}_e(C_2) = 1$, $\mathrm{Var}_e(X) = 3 + w$, and $\Cov_e(C_2, X) = -1$, so the two covariate distributions are identical. Repeating the risk calculation with $\mathrm{Var}_e(C_1) = 1 + w$ in $e_A$ and $q_e = 1 + w$ in $e_B$ gives
\begin{align*}
\Delta_{e_A} &= \beta_3\big[(2-\beta_3)(1 + w + \mu_0^2) - \beta_3\big] > 0,\\
\Delta_{e_B} &= \beta_3\big[(2-\beta_3)(1 + \mu_0^2) - \beta_3(1 + w)\big],
\end{align*}
so the proxy should be kept in $e_A$ for every $w$ and dropped in $e_B$ once $w > (2-\beta_3)(1 + \mu_0^2)/\beta_3 - 1$.

\section{Proofs}
\label{sec:s_proofs}

Throughout, $\mathcal{F}$ is the hypothesis class of \Cref{subsec:theorem}, and suprema over $\mathcal{F}$ are assumed measurable.

\begin{lem}[Uniform convergence]
\label{lem:uniform_conv}
For any $\delta \in (0,1)$, with probability at least $1-\delta$,
\[
U := \sup_{\mathcal{F}}\big| \widehat{\mathcal{S}}(g) - \mathcal{S}(g) \big| \;\le\; 2 r_m + B\sqrt{\frac{\log(1/\delta)}{2m}} + a_n .
\]
\end{lem}

\begin{proof}
Split $\widehat{\mathcal{S}}(g) - \mathcal{S}(g)$ into $\widehat{\mathcal{S}}(g) - \mathbb{E}\,\widehat{\mathcal{S}}(g)$ and $\mathbb{E}\,\widehat{\mathcal{S}}(g) - \mathcal{S}(g)$.

For the first part, $\widehat{\mathcal{S}}(g)$ averages $m$ i.i.d.\ terms $c_j(g) = \sum_z g_z(u_{e_j})\, \widehat R_{e_j}(z) \in [0, B]$, one per training environment. Changing one environment changes the supremum over $\mathcal{F}$ by at most $B/m$, so McDiarmid's inequality and symmetrization \citep{bartlett2002rademacher} give, with probability at least $1-\delta$,
\[
\sup_{\mathcal{F}} \big| \widehat{\mathcal{S}}(g) - \mathbb{E}\,\widehat{\mathcal{S}}(g) \big| \le 2 r_m + B\sqrt{\frac{\log(1/\delta)}{2m}},
\qquad
r_m = \mathbb{E} \sup_{\mathcal{F}} \Big| \frac{1}{m}\sum_{j=1}^m \xi_j\, c_j(g) \Big|,
\]
where $\xi_1, \dots, \xi_m$ are independent random signs.

For the second part, a training environment and its representation have the same joint distribution as the test environment and its representation, because both representations are computed from $n$ observations. Since the weights $g_z(u_e)$ are nonnegative and sum to one,
\[
\big| \mathbb{E}\,\widehat{\mathcal{S}}(g) - \mathcal{S}(g) \big|
= \Big| \mathbb{E} \sum_z g_z(u_e) \big\{ \widehat R_e(z) - R_e(z) \big\} \Big|
\le \mathbb{E} \sup_{\mathcal{F}} \max_z \big| \widehat R_e(z) - R_e(z) \big| = a_n .
\]
\end{proof}

\begin{proof}[Proof of \Cref{theorem:finite}]
On the event of \Cref{lem:uniform_conv}, for every $g \in \widehat{\mathcal{G}}$,
\[
\mathcal{S}(\widehat g)
\le \widehat{\mathcal{S}}(\widehat g) + U
\le \widehat{\mathcal{S}}(g) + \varepsilon_{\mathrm{opt}} + U
\le \mathcal{S}(g) + \varepsilon_{\mathrm{opt}} + 2U .
\]
The first and last inequalities use the definition of $U$, and the middle one uses the condition on $\widehat g$. Taking the infimum over $g \in \widehat{\mathcal{G}}$ and bounding $U$ by \Cref{lem:uniform_conv} gives the theorem.
\end{proof}

\begin{proof}[Proof of \Cref{theorem:asymptotic}]
Let $\Delta = \mathcal{S}(\widehat g) - \inf_{g \in \widehat{\mathcal{G}}} \mathcal{S}(g)$, which is nonnegative because $\widehat g \in \widehat{\mathcal{G}}$. Fix $\delta \in (0,1)$ and $\eta > 0$. By \Cref{theorem:finite}, $\Delta$ exceeds its bound with probability at most $\delta$, and the bound tends to zero in probability. Hence $\Pr(\Delta > \eta) \le \delta + o(1)$. Since $\delta$ is arbitrary, $\Pr(\Delta > \eta) \to 0$.
\end{proof}

\begin{proof}[Proof of \Cref{prop:linear_scm}]
In an acyclic linear structural causal model, solving the structural equations writes each variable as a linear function of the noise terms and the interventions. Let $t_k \in \mathbb{R}^p$ and $s_k \in \mathbb{R}$ denote the total causal effects of variable $k$ on $W$ and on $Y$, so that $t_k$ determines the trace of an intervention on $k$. If environment $e$ shifts the mean of each variable $k$ by $\nu_{e,k}$ and adds independent mean-zero noise with variance $\omega_{e,k}$, then
\[
\mu_{W,e} = \mu_{W,0} + \sum_k \nu_{e,k}\, t_k,
\qquad
\mu_{Y,e} = \mu_{Y,0} + \sum_k \nu_{e,k}\, s_k,
\]
\[
\Sigma_{W,e} = \Sigma_0 + \sum_k \omega_{e,k}\, t_k t_k^\top,
\qquad
\Cov_e(W, Y) = \gamma_0 + \sum_k \omega_{e,k}\, s_k t_k,
\]
where $\mu_{W,0}$, $\mu_{Y,0}$, $\Sigma_0$, and $\gamma_0$ do not depend on $e$. Condition (i) states that whenever two mean-shift patterns $\nu$ and $\nu'$ satisfy $\sum_k \nu_k t_k = \sum_k \nu'_k t_k$, they also satisfy $\sum_k \nu_k s_k = \sum_k \nu'_k s_k$. Hence any two environments with the same $\mu_{W,e}$ have the same $\mu_{Y,e}$. Condition (ii) states that whenever two noise patterns $\omega$ and $\omega'$ satisfy $\sum_k \omega_k t_k t_k^\top = \sum_k \omega'_k t_k t_k^\top$, they also satisfy $\sum_k \omega_k s_k t_k = \sum_k \omega'_k s_k t_k$. Hence any two environments with the same $\Sigma_{W,e}$ have the same $\Cov_e(W, Y)$, so $\Cov_e(W, Y)$ is determined by $\Sigma_{W,e}$. Since $\mu_{Y,e}$ is determined by $\mu_{W,e}$, every term in the expansion of $R_e(z) - \Var_e(Y)$ in \Cref{subsec:linear_scm} is determined by $\mu_{W,e}$ and $\Sigma_{W,e}$. 
\end{proof}

\section{Additional Simulation Results for EACS}
\label{sec:s_sim_additional}

This section gives the full setup and results for the running-example simulation study summarized in \Cref{subsec:sim33}, followed by two related studies in \Cref{sec:s_contextual_all,sec:s_ambiguity}.

\noindent\textbf{Setup.}
The library $Z$ consists of the four subsets of $\{C_2, X\}$, each fitted by pooled least squares with an intercept on all training observations. Each environment is summarized by $u_e = (r_e, s_{2,e}, s_{3,e})$: the correlation between $C_2$ and $X$ and their standard deviations. The gate is a small decision tree fitted to minimize the selection risk \eqref{eq:selection-risk}, and it selects a single subset for each environment. The tree has at most two levels of splits, each split compares one summary statistic with a threshold chosen from the deciles of its training values, and the tree is found by exhaustive search over these choices. We compare EACS with two fixed predictors from the same library: the causal-only predictor using $\{C_2\}$ and the proxy-augmented predictor using $\{C_2, X\}$.

\noindent\textbf{Experimental conditions.}
We vary one factor at a time: (i) the number of training environments per shift type, taking values $5$, $20$, and $100$; (ii) the number of observations per training environment, from $100$ down to $5$; (iii) the summary, using the full $u_e$ or a single statistic; and (iv) the largest perturbation level used in training, from $4$ down to $0$. Unless varied, there are $20$ training environments per shift type with $100$ observations each, training outcome-noise SD $1$, largest perturbation level $4$, and the full summary. Within each shift type, the training environments use equally spaced perturbation levels between $0$ and this largest level. For testing, each of the three shift types uses $20$ equally spaced perturbation levels between $0$ and $4$, giving $60$ test environments with $100$ observations each and outcome noise $1$. For factors (i) and (ii), we additionally vary the training outcome-noise SD $\sigma_Y \in \{1, 5, 10\}$. Results are averaged over 1{,}000 replications.

\begin{table}[t]
\centering
\caption{Mean test MSE in the running example over the $60$ test environments. Each row changes the stated factors from the reference setting: $20$ training environments per shift type with $100$ observations each, training outcome-noise SD $\sigma_Y = 1$, training perturbations up to $4$, and the full summary.}
\label{tab:sim_summary}
\begin{tabular}{llccc}
\toprule
Factor & Setting & EACS & $\{C_2\}$ & $\{C_2, X\}$ \\
\midrule
Reference & & 2.11 & 3.38 & 2.29 \\
\addlinespace
Training environments & 5 per shift type, $\sigma_Y = 1$ & 2.15 & 3.38 & 2.30 \\
 & 5 per shift type, $\sigma_Y = 10$ & 2.80 & 3.52 & 2.49 \\
Observations & 5 per environment, $\sigma_Y = 1$ & 2.20 & 3.40 & 2.32 \\
 & 5 per environment, $\sigma_Y = 10$ & 3.96 & 4.08 & 3.34 \\
Summary & SD of $C_2$ only & 2.36 & 3.38 & 2.29 \\
 & SD of $X$ only & 2.21 & 3.38 & 2.29 \\
 & Correlation only & 2.14 & 3.38 & 2.29 \\
Training perturbations & up to 2 & 2.10 & 3.49 & 2.33 \\
 & none & 2.42 & 3.83 & 2.42 \\
\bottomrule
\end{tabular}
\end{table}

\begin{figure}[t]
    \centering
    \includegraphics[width=\linewidth]{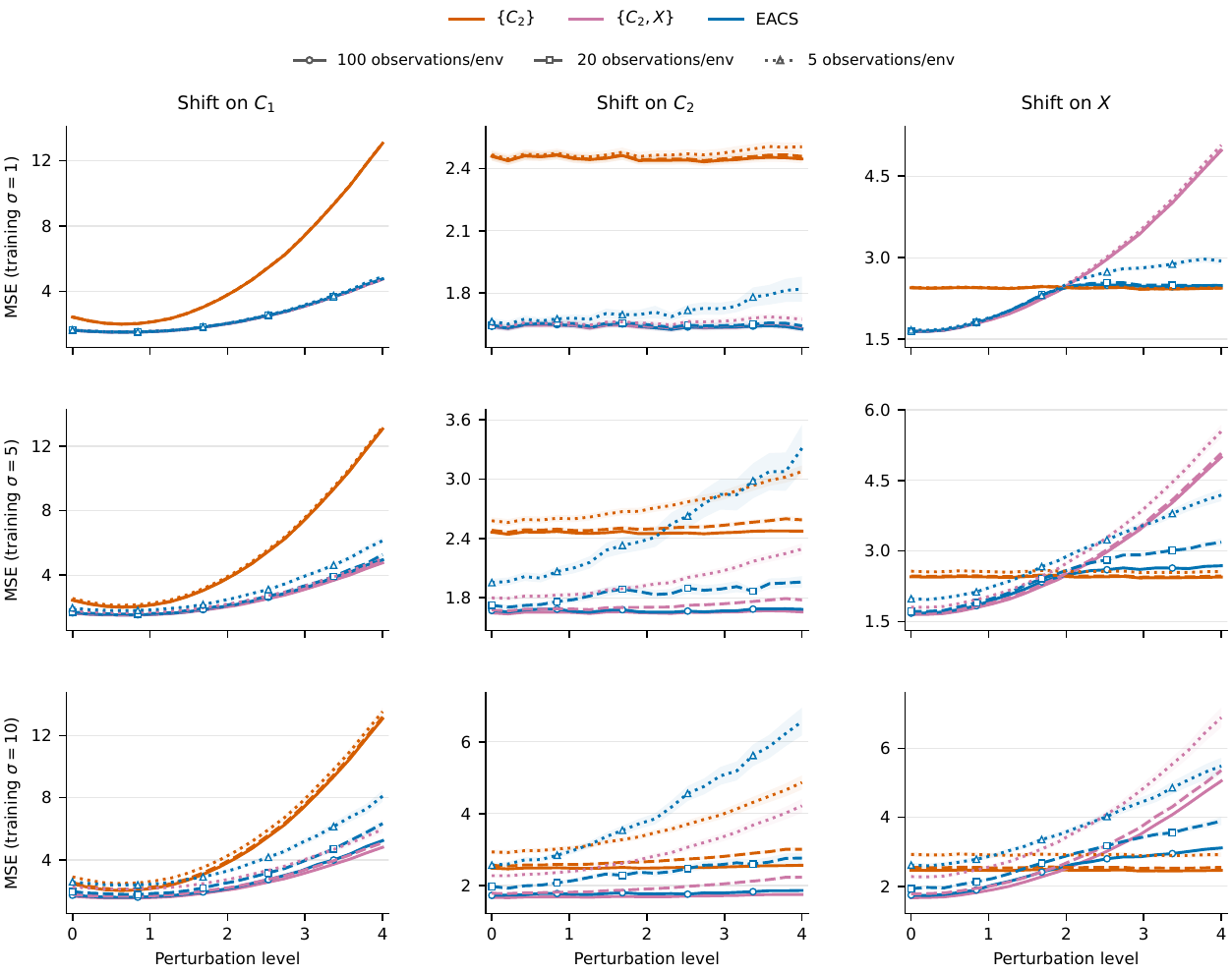}
\caption{MSE for varying observations per training environment and training outcome noise. Colors identify predictors; line styles and markers identify the number of observations per environment. Shaded regions show pointwise 95\% confidence intervals over 1{,}000 replications.\label{fig:sim_a_sam_mse}}
\end{figure}

\paragraph{Training environments and observations.}
The prediction error of EACS decreases with more training environments, more observations per environment, and lower training outcome noise (\Cref{fig:sim_a_env_mse,fig:sim_a_sam_mse}). When outcome noise is low, EACS has the lowest prediction error even with few training environments or observations; when outcome noise is high, it can be worse than always keeping the proxy (\Cref{tab:sim_summary}).

\begin{figure}[!htbp]
  \centering
  \includegraphics[width=1\linewidth]{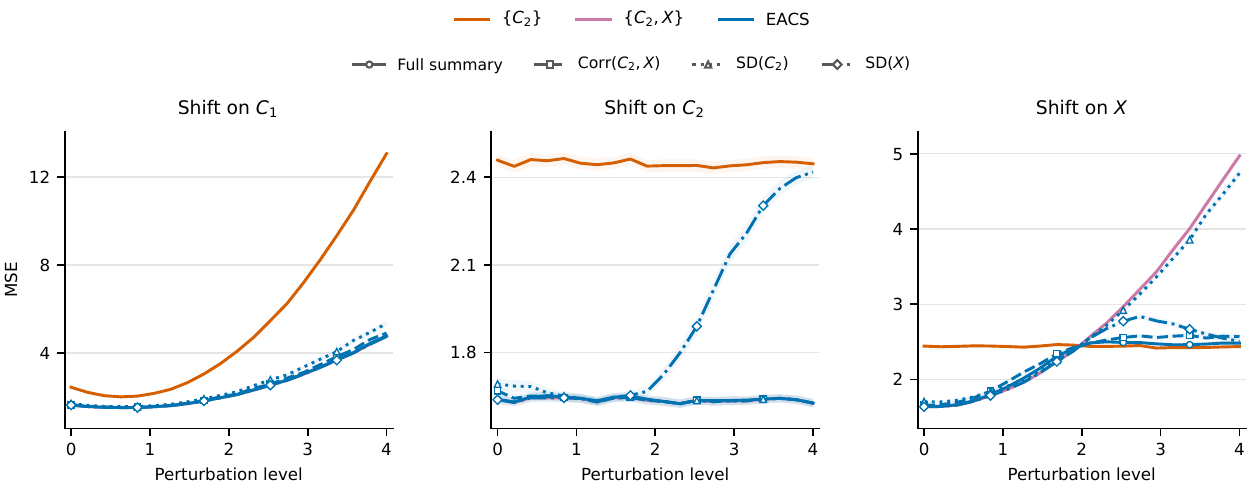}
\caption{MSE for different summaries. Shaded regions show pointwise 95\% confidence intervals over 1{,}000 replications.}
\label{fig:supp_sta_mse}
\end{figure}

\begin{figure}[!htbp]
  \centering
  \includegraphics[width=\linewidth]{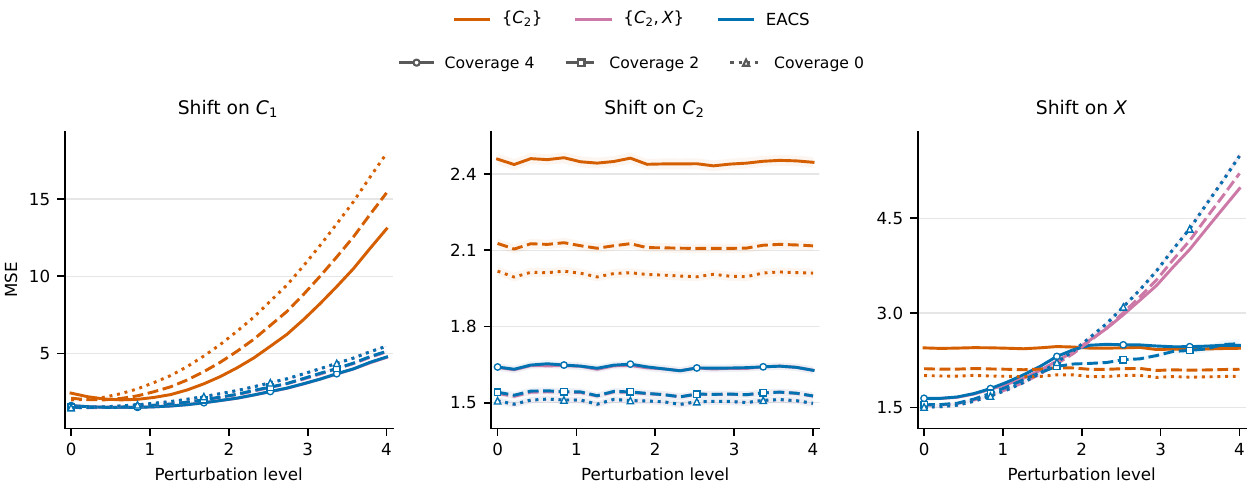}
\caption{MSE for varying training coverage. Line styles and markers identify the maximum training perturbation. Shaded regions show pointwise 95\% confidence intervals over 1{,}000 replications.}
\label{fig:supp_lev_mse}
\end{figure}

\paragraph{Summary informativeness and perturbation coverage.}
The full summary yields the lowest error, and single statistics are less informative (\Cref{fig:supp_sta_mse}). EACS remains accurate when the training perturbations are weaker than those at test time, but it cannot learn when to drop the proxy if no training environment is perturbed (\Cref{fig:supp_lev_mse}).

\subsection{Comparison with contextual prediction: all factors}
\label{sec:s_contextual_all}

\begin{figure}[t]
\centering
\includegraphics[width=\linewidth]{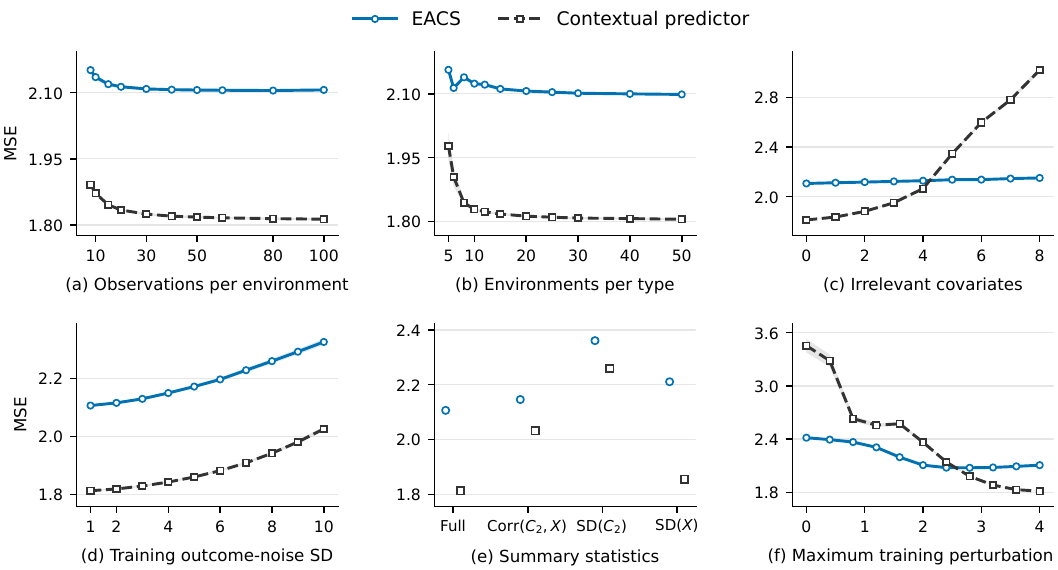}
\caption{EACS versus the contextual predictor in the running example, for all six factors. Each panel varies one factor from the reference setting: (a) observations per training environment, (b) training environments per shift type, (c) irrelevant covariates, (d) training outcome-noise SD, (e) the summary, either full or a single statistic, and (f) the largest training perturbation. Points show mean test MSE over 1{,}000 replications, with pointwise 95\% confidence intervals.}
\label{fig:contextual_all}
\end{figure}

\Cref{fig:contextual_all} shows all six factors varied in \Cref{sec:contextual_main}. Besides the number of irrelevant covariates and the range of training perturbations, we vary the number of observations per environment (from $8$ to $100$), the number of training environments per shift type (from $5$ to $50$), the training outcome-noise SD (from $1$ to $10$), and the summary, comparing the full summary with summaries that use only the correlation between $C_2$ and $X$, only the SD of $C_2$, or only the SD of $X$, as in \Cref{fig:supp_sta_mse}. None of these four variations gives EACS an advantage: with no irrelevant covariates and training shifts that cover the test shifts, neither cost of contextual prediction in \Cref{rem:env_specific} arises, and the contextual predictor remains more accurate throughout.

\subsection{Shifts that leave the same trace}
\label{sec:s_ambiguity}

This section simulates the two pairs of shifts from \Cref{subsec:linear_scm} that leave the same trace. Each experiment uses one pair: noise added to $C_1$ or to $X$, or a mean shift on $C_1$ or on $X$. Otherwise, it follows the reference setting of this appendix and uses the same library and gate, fitted separately for each experiment. The summary also includes the means of $C_2$ and $X$, so that mean shifts are visible to the gate. 

\begin{figure}[t]
\centering
\includegraphics[width=\linewidth]{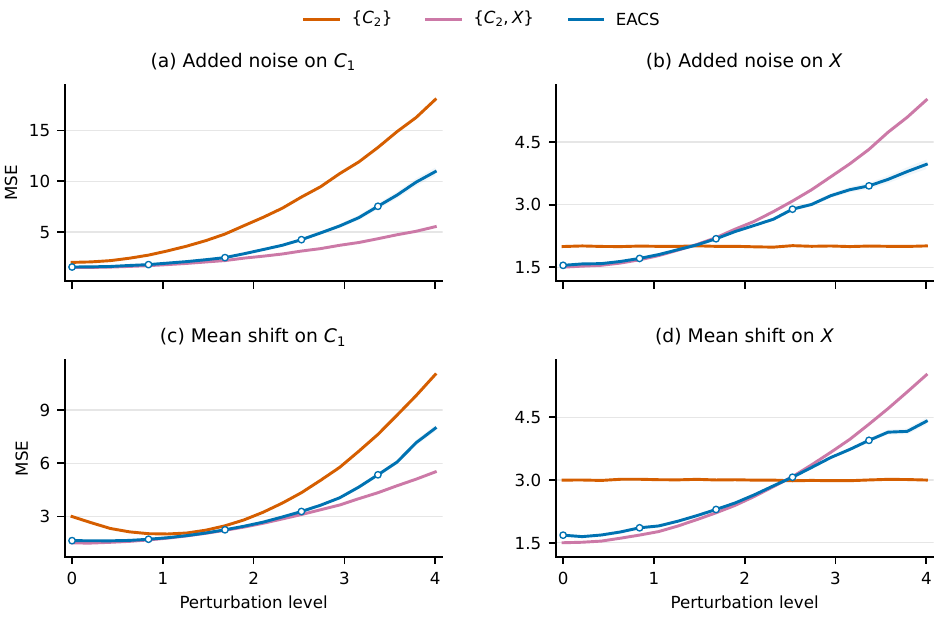}
\caption{MSE of EACS and of the fixed predictors using $\{C_2\}$ or $\{C_2, X\}$ under two pairs of shifts that leave the same trace: noise added to $C_1$ (a) or to $X$ (b), and a mean shift on $C_1$ (c) or on $X$ (d). Each row is a separate experiment with its own training environments. Shaded regions show pointwise 95\% confidence intervals over 1{,}000 replications.}
\label{fig:ambiguity}
\end{figure}

\Cref{fig:ambiguity} shows the results over $1{,}000$ replications. In both experiments, keeping $X$ is better under the shift on $C_1$, and dropping it is better under large shifts on $X$. Because the covariates have the same distribution under both shifts, EACS chooses each subset with the same probability under both, so its error lies between those of the two fixed predictors. 

\section{Prior Causal Knowledge in EACS: Full Algorithms, Theory, and Simulations}
\label{sec:s_causal_full}

Causal knowledge is often available from domain expertise or prior causal analysis \citep{peters2016causal, heinze2018invariant, Fan2024environment, Wu2025bayesian, hoyer2009additive, buhlmann2014cam, shimizu2006lingam, peters2014identifiability}. Incorporating it requires the selector to retain the known causal parents while still allowing it to include non-causal proxies when they improve prediction under the observed shift.

\subsection{Algorithms}

\Cref{alg:eafs1-causal,alg:eafs2-causal} are the constrained versions of \Cref{alg:eafs1,alg:eafs2}. The discrete version restricts the library to $Z_S$, and the continuous version fixes the feature weights of the covariates in $S$ at one. All other steps are unchanged.

\begin{algorithm}[t]
\caption{EACS (discrete selector) with causal constraints}
\label{alg:eafs1-causal}
\textbf{Input:} The inputs of \Cref{alg:eafs1} and the indices $S$ of the covariates required in every candidate subset.
\begin{algorithmic}[1]
\State Restrict the library to $Z_S = \{z \in Z : z_j = 1 \text{ for all } j \in S\}$.
\State Run \Cref{alg:eafs1} with $Z_S$ in place of $Z$, so that the gate assigns weights only to subsets in $Z_S$.
\end{algorithmic}
\end{algorithm}

\begin{algorithm}[t]
\caption{EACS via continuous feature gating with causal constraints}
\label{alg:eafs2-causal}
\textbf{Input:} The inputs of \Cref{alg:eafs2} and the indices $S$ of the known causal covariates.
\begin{algorithmic}[1]
\Statex \textbf{Training:}
\State Initialize $\vartheta = (\theta_p, \theta_{\mathrm{gate}}, \theta_{\mathrm{env}})$.
\State For each $e \in \mathcal{E}_{\mathrm{train}}$, compute $u_e = f_{\mathrm{env}}(\{W_{i,e}\}_{i=1}^{n_e}; \theta_{\mathrm{env}})$ and $\alpha_e = f_{\mathrm{gate}}(u_e; \theta_{\mathrm{gate}})$, and form the feature weights
\[
\tilde z_{e,j} =
\begin{cases}
  1, & j \in S,\\
  \sigma(\alpha_{e,j}/\tau), & j \notin S.
\end{cases}
\]
\State Minimize the environment-averaged loss of \Cref{alg:eafs2} jointly over $\vartheta$.
\medskip
\Statex \textbf{Test time:} given a new environment $e^*$ with unlabeled covariates, compute $u_{e^*}$ and $\tilde z_{e^*}$ in the same way, with $\tilde z_{e^*,j} = 1$ for $j \in S$, and predict $\widehat Y_{i,e^*} = p_{\hat\theta_p}(\tilde z_{e^*} \circ W_{i,e^*})$.
\end{algorithmic}
\end{algorithm}

\subsection{Theory}

The constrained discrete selector is \Cref{alg:eafs1} applied to the library $Z_S$, so \Cref{theorem:finite,theorem:asymptotic} hold with $Z$ replaced by $Z_S$ throughout: the selection risk, the gate class, and the hypothesis class $\mathcal{F}$ of \Cref{sec:s_proofs} are all restricted to $Z_S$, and the proofs are unchanged. Because $r_m$ and $a_n$ are suprema over the class and the restricted class is contained in the unrestricted one, both can only decrease. 

\subsection{Running example continued: evaluating EACS with causal constraints}
\label{subsec:s_causal_sim}

We repeat the simulations of \Cref{sec:s_sim_additional} with the constraint that $C_2$ appears in every subset, which reduces the library to $\{C_2\}$ and $\{C_2, X\}$. The gate class, summaries, and experimental conditions are unchanged.

\begin{figure}[t]
    \centering
    \includegraphics[width=\linewidth]{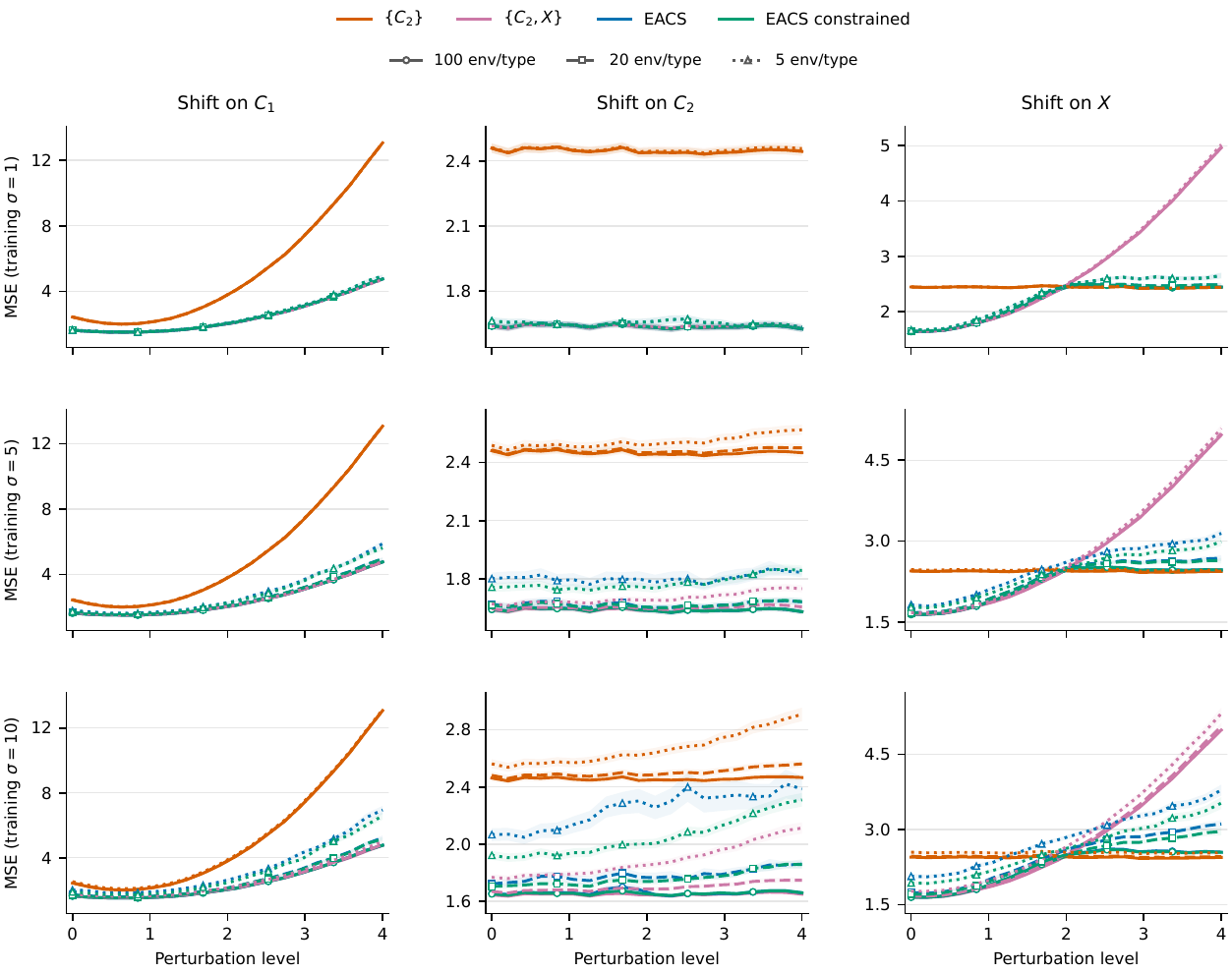}
\caption{MSE of constrained and unconstrained EACS for varying numbers of training environments and training outcome noise. Shaded regions show pointwise 95\% confidence intervals over 1{,}000 replications.}
\label{fig:sim_a_env_cau_mse}
\end{figure}

\begin{figure}[t]
    \centering
    \includegraphics[width=\linewidth]{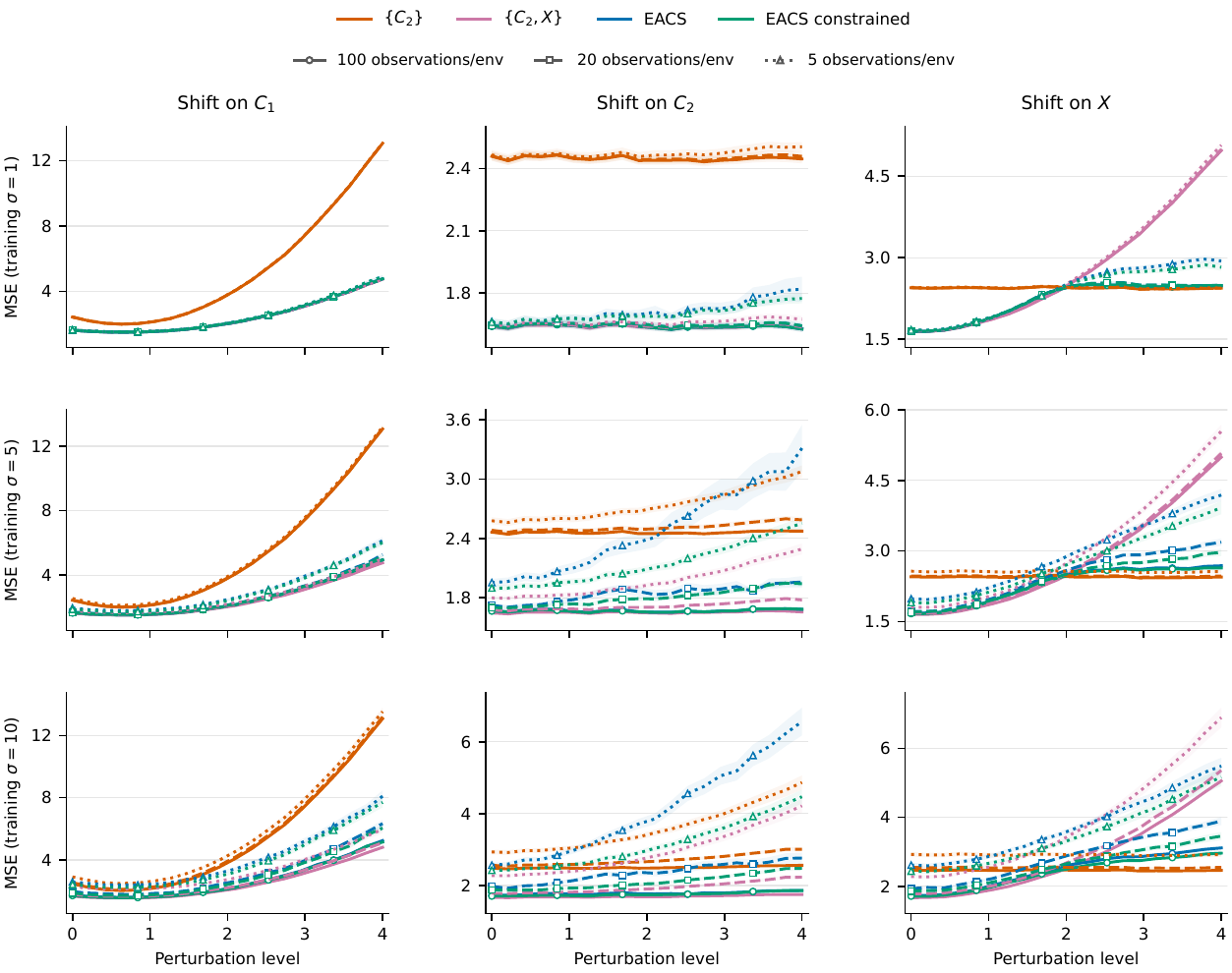}
\caption{MSE of constrained and unconstrained EACS for varying observations per training environment and training outcome noise. Shaded regions show pointwise 95\% confidence intervals over 1{,}000 replications.}
\label{fig:sim_a_sam_cau_mse}
\end{figure}

\begin{figure}[t]
  \centering
  \includegraphics[width=\linewidth]{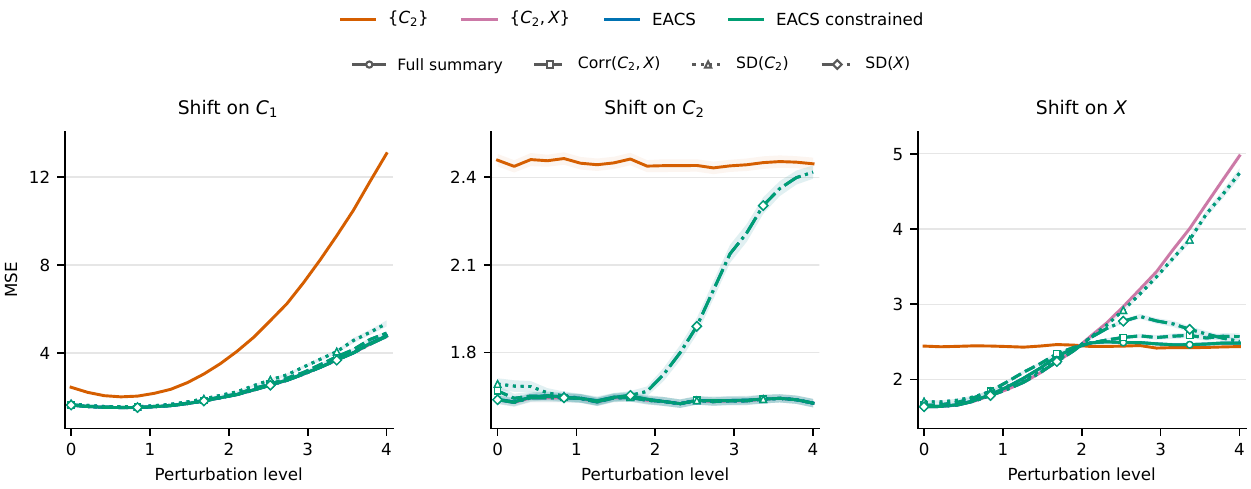}
\caption{MSE of constrained and unconstrained EACS for different summaries. Shaded regions show pointwise 95\% confidence intervals over 1{,}000 replications.}
\label{fig:supp_sta_cau_mse}
\end{figure}

\begin{figure}[t]
  \centering
  \includegraphics[width=\linewidth]{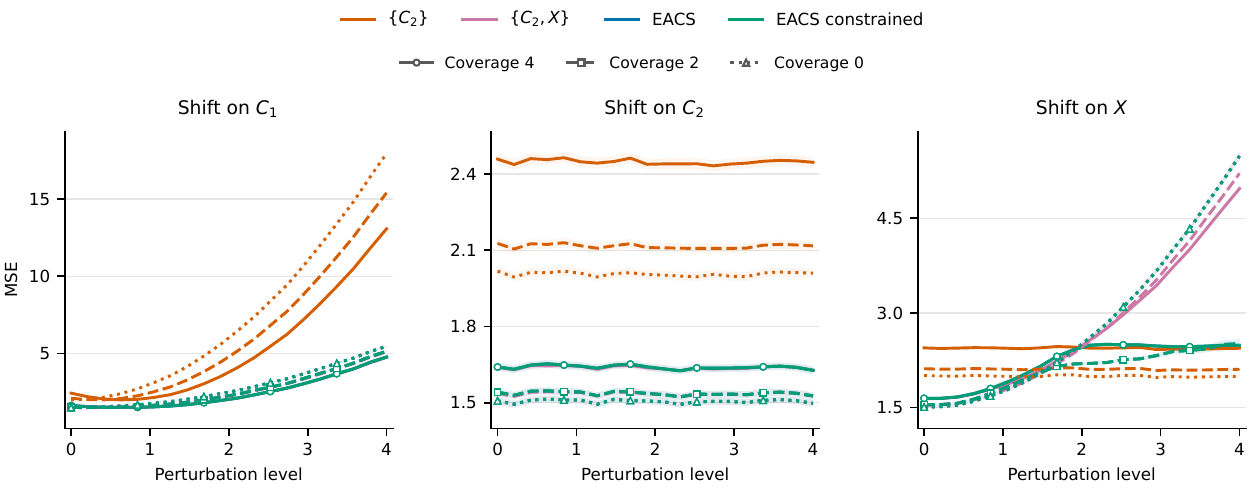}
\caption{MSE of constrained and unconstrained EACS for varying training coverage. Shaded regions show pointwise 95\% confidence intervals over 1{,}000 replications.}
\label{fig:supp_lev_cau_mse}
\end{figure}

The constraint lowers prediction error when outcome noise is high and either the training environments or the observations per environment are few (\Cref{fig:sim_a_env_cau_mse,fig:sim_a_sam_cau_mse}). Otherwise, the constrained and unconstrained selectors perform similarly. In particular, the constraint does not compensate for an uninformative summary or for limited perturbation coverage during training (\Cref{fig:supp_sta_cau_mse,fig:supp_lev_cau_mse}).

\section{Implementation Details for the Data Applications}
\label{sec:implementation}

\noindent\textbf{Shared protocol.}
An environment is a calendar day for bike sharing and a state for ACS Income. The environments are split into five contiguous blocks; each block is held out in turn, and all tuning uses three-fold cross-validation (CV) over whole training environments, after which the selected configuration is refitted on all training environments. All preprocessing is fitted on the training environments only. The reported mean is the average MSE over all held-out environments, each weighted equally, and the reported SD is the SD of the five fold means. Methods with random initialization or random features are run with three seeds, and their test losses are averaged over seeds.

\subsection{Bike-sharing dataset}

\noindent\textbf{Data and response.}
We use the hourly data for 2011--2012 \citep{fanaee2013bike, fanaee2013event}: $17{,}378$ observations on the $730$ days with at least three hourly records, with the weather covariates temperature, feeling temperature, humidity, and wind speed. The response is the square root of the rental count minus its holiday and weekday effects, which are estimated on the training days and applied to all days. This adjustment is shared by all methods. The weather covariates are standardized using the training days.

\noindent\textbf{Static baselines.}
OLS is least squares on all four weather covariates. The lasso penalty $\lambda$ is selected from $\{10^{-3}, 10^{-2}, 10^{-1}, 1\}$. Anchor regression \citep{Rothenhausler2021anchor} uses day indicators as anchors, with $\gamma$ selected from $\{0.2, 0.5, 1, 2, 5, 10, 20\}$. ICP \citep{peters2016causal} tests each covariate subset for invariance across days, comparing the residual mean and variance of each day with those of the remaining days at significance level $0.05$, with a Bonferroni correction. CV selects $\lambda = 10^{-2}$ and $\gamma = 1$ in every fold, and ICP accepts no subset in any fold.

\noindent\textbf{Summary.}
Each day is summarized by the means, SDs, and pairwise partial correlations of the four standardized covariates, a $14$-dimensional vector that is standardized across training days.

\noindent\textbf{EACS.}
The library contains least-squares fits on each of the $16$ subsets of the weather covariates, including the intercept-only model. We consider three gate families: linear softmax models, small ReLU networks, and random forests of $100$ trees. The networks have one hidden layer of $32$ units or two of $64$ and $32$ units; the linear and network gates use learning rate $3 \times 10^{-4}$ or $10^{-3}$ and $500$ or $2{,}000$ training steps. The forests have depth $3$, $5$, or $8$ and minimum leaf size $10$ or $30$. All are trained to minimize the selection risk \eqref{eq:selection-risk}. CV selects the gate family, its hyperparameters, and whether to use hard selection or soft averaging. The random forest with soft averaging is selected in every fold.

\noindent\textbf{Contextual predictor.}
The contextual predictor of \Cref{sec:applications} uses the same summary as EACS and is fitted by ridge regression, with the penalty selected from $11$ logarithmically spaced values between $10^{-6}$ and $10^{4}$.

\subsection{ACS Income dataset}

\noindent\textbf{Data and covariates.}
We use $1{,}664{,}500$ records from the 2018 ACS Income data \citep{ding2021retiring}, prepared following \citet{jeong2025out}, with log income as the response. The covariates are age and weekly working hours, which are numeric, and class of worker, education, marital status, occupation, relationship, sex, and race, which are categorical. Each categorical variable is one-hot encoded with one reference level, giving $590$ covariate columns. The two numeric covariates are standardized using the training states.

\noindent\textbf{Static baselines.}
OLS is least squares on all covariates. The lasso penalty $\lambda$ is selected from $11$ logarithmically spaced values between $10^{-5}$ and $1$. Anchor regression uses state indicators as anchors, with $\gamma$ selected from $\{0.2, 0.5, 1, 2, 5, 10, 20\}$. CV selects $\lambda = 10^{-5}$ and $\gamma = 0.5$ in every fold.

\noindent\textbf{EACS.}
We use the continuous version of \Cref{alg:eafs2} with a shared linear predictor and one feature weight per covariate column. The gate is a small ReLU network. The summary $u_e$ is a DeepSets encoding \citep{zaheer2017deep}: each observation is embedded in $16$ dimensions by a one-layer ReLU network, and $u_e$ is the average embedding over the state. The encoder, gate, and linear predictor are trained jointly to minimize the prediction error.

\noindent\textbf{Contextual predictor.}
The contextual predictor has the linear form of \Cref{sec:applications}, but its summary $u_e$ is learned rather than fixed: it uses its own DeepSets encoder, with the same architecture as in EACS, trained jointly with the regression coefficients in the same way.

\noindent\textbf{Optimization and tuning.}
EACS and the contextual predictor are trained with Adam, with learning rate $10^{-3}$ and mini-batches of four states. For each of them, CV selects an $\ell_2$ penalty from $\{0, 10^{-4}, 10^{-2}\}$ and the number of epochs from $\{25, 100, 400, 1{,}600\}$, and the selected model is refitted on all training states.

\noindent\textbf{Fixed marginal summary.}
\Cref{tab:acs_summary} compares the learned summary with a fixed one: the means and SDs of the numeric covariates and the proportions of each category level. EACS uses this summary as the input to its gate, and the contextual predictor is fitted by ridge regression, with the penalty selected from the same grid as in bike sharing. EACS is then slightly less accurate than with the learned summary, and the contextual predictor is much less accurate, mainly because of Alaska, whose summary lies far outside the range of the training states.

\begin{table}[t]
\centering
\caption{ACS Income: learned and fixed marginal summaries. Mean MSE over all held-out states and SD of the five fold means.}
\label{tab:acs_summary}
\begin{tabular}{lcc}
\toprule
Method & Mean MSE & Fold SD \\
\midrule
EACS (learned summary) & 0.6140 & 0.0205 \\
EACS (fixed marginal summary) & 0.6152 & 0.0221 \\
Contextual (learned summary) & 0.6134 & 0.0196 \\
Contextual (fixed marginal summary) & 1.1792 & 0.8998 \\
\bottomrule
\end{tabular}
\end{table}

\bibliography{ref}

\end{document}